\documentclass[acmsmall, screen]{acmart}
\usepackage{booktabs}
\usepackage[linesnumbered]{algorithm2e}

\begin{document}
\title{HW/SW Framework for Improving the Safety of Implantable and Wearable Medical Devices}

%

\author{Malin Prematilake}
\email{mpremati@purdue.edu}
\affiliation{%
  \institution{Purdue University, USA}
  \streetaddress{School of ECE }
  \city{West Lafayette}
  \state{Indiana}
  \postcode{47907}
}
\author{Younghyun Kim}
\email{younghyun.kim@wisc.edu}
\affiliation{%
  \institution{University of Wisconsin\-Madison, USA}
  \streetaddress{School of ECE }
  \city{West Lafayette}
  \state{Indiana}
  \postcode{47907}
}
\author{Vijay Raghunathan}
\email{vr@purdue.edu}
\author{Anand Raghunathan}
\email{raghunathan@purdue.edu}
\affiliation{%
  \institution{Purdue University, USA}
  \streetaddress{School of ECE }
  \city{West Lafayette}
  \state{Indiana}
  \postcode{47907}
}
\author{N.K. Jha}
\email{jha@princeton.edu}
\affiliation{%
  \institution{Princeton University, USA}
  \streetaddress{School of ECE }
  \city{West Lafayette}
  \state{Indiana}
  \postcode{47907}
}

%
\begin{abstract}
Implantable and wearable medical devices (IWMDs) are widely used for the monitoring and therapy of an increasing
range of medical conditions. Improvements in medical devices, enabled by advances in low-power processors,
more complex firmware, and wireless connectivity, have greatly improved therapeutic outcomes and patients' quality-of-life.
However, security attacks, malfunctions and sometimes user errors have raised great concerns regarding the safety of IWMDs. In this work, we
present a HW/SW (Hardware/Software) framework for improving the safety of IWMDs,  wherein a
set of safety rules and a rule check mechanism are used to monitor both the
extrinsic state (the patient's physiological parameters sensed by the IWMD) and the internal state of the IWMD
(I/O activities of the microcontroller) to infer unsafe operations that may be triggered by user errors,
software bugs, or security attacks. We discuss how this approach can be realized in the context of a artificial pancreas with wireless connectivity and implement a prototype to demonstrate its effectiveness in improving safety at modest overheads.
\end{abstract}

\ccsdesc[500]{Applied computing~Consumer health}
\ccsdesc[500]{Computer systems organization~Firmware}
\ccsdesc[500]{Computer systems organization~Embedded hardware}
\ccsdesc[300]{Security and privacy~Embedded systems security}

%
\keywords{implantable and wearable medical devices, framework, architecture, safety, safety rules, coprocessor, artificial pancreas}

\maketitle

\section{Introduction}\label{sec:introduction}
The development and adoption of implantable
and wearable medical devices (IWMDs) for personal health care has greatly improved therapeutic outcomes and quality-of-life for patients by enabling continuous and autonomous management of medical
conditions such as diabetes and cardiac arrhythmia. Much like other computing platforms, IWMDs have
gone through a rapid evolution, greatly benefiting from advances in low-power processing
and wireless communication technologies. Modern IWMDs are equipped with a microcontroller 
that can execute sophisticated algorithms for detection and response in real-time. They
also provide wireless connectivity for post-deployment diagnostics, tuning of therapy,
and access to medical data for the patient and healthcare provider.

Unfortunately, these advances are accompanied by higher hardware and software complexity, leading to 
an increase in unreliability as well as security vulnerabilities, which lead to
safety risks to the patient~\cite{Halperin-SSP08,Li-Healthcom11,Radcliffe-BlackHat11,Alemzadeh-SP13,Zhang-ProcIEEE14,Kim-IBM15,Z-2544-2014,FDA14,Ray-DnT15,Rushanan-SP14}. Complex IWMD firmware often contains bugs that
cannot be completely eliminated at design time. 
A recent research study attributed 64.3\% of medical device recalls issued by the U.S.
Food and Drug Administration (FDA) between 2006 and 2011 to software-related issues~\cite{Alemzadeh-SP13}.
Researchers have also demonstrated how wireless connectivity can be exploited to remotely
trigger adverse activities in IWMDs~\cite{Halperin-SSP08,Li-Healthcom11,Radcliffe-BlackHat11}.

We observe that regardless of cause --- firmware bug, malicious attack, or even user error ---
unsafe operation of IWMDs is the utmost concern of patients since it can adversely affect
their health or even be life-threatening. While there are other valid concerns
such as privacy, we focus our effort on preventing unsafe operations. Broadly speaking,
we address this challenge by enhancing the IWMD's HW/SW (Hardware/Software) so that it is able to identify
unsafe IWMD operations as soon as possible and even prevent them before they can have an effect on the patient.
There are three main challenges involved in designing such a safety mechanism for IWMDs. 
First, deciding whether a given IWMD behavior is safe or unsafe is context-dependent. For example,
in the case of an insulin pump, a bolus insulin dose can be determined to be safe or unsafe
depending on the patient's blood glucose (BG) level; similarly, a high-voltage defibrillation pulse generated
by an implantable cardioverter defibrillator (ICD) can be deemed unsafe if it is produced shortly after a wireless packet is
processed and the sensed heartbeat is normal. Thus, high-level context awareness is
essential for accurate decision making regarding safety.
Second, unsafe operations should be identified and blocked in a proactive manner before
they are actually performed because operations of IWMDs may be irreversible once they
take place (e.g., infused insulin cannot be removed).
Third, it is desirable that the safety checking be performed in a
computing environment
that is isolated from the normal functions of the IWMD to prevent contamination
from failures and security attacks. This will also facilitate integration into existing
IWMDs that are designed under extremely conservative development and regulatory processes by minimizing HW/SW modification.
Further, due to the stringent power constraints imposed on IWMDs, the power
overheads imposed by the checking mechanism should kept be as low as possible.

In this paper, we propose a \textit{safety framework} that addresses the challenges explained above. 
The framework consists of two parts: a set of rules that guarantees the safety of the patient and a \textit{rule-check} mechanism that checks
whether these rules are satisfied by the IWMD.
Based on the nature of the rules, some of them can be validated during the development stage of the firmware (i.e. static rule-check) while the rest can 
be validated during runtime (i.e. dynamic rule-check). An existing source code analyzer can be used to accomplish the static rule-check. However, the 
mechanism used to conduct a validation during runtime should adhere to the following: It should operate completely independent from the regular IWMD 
operations, it should have minimum effect on the operations of the IWMD although it should be able to prevent unsafe operations. Based on these
two notions, we decided to use a \textit{safety coprocessor} for dynamic rule checking.

While the concept of a coprocessor for reliability~\cite{Hossain-CODES12,Nakka-DSN04}, security~\cite{IBM4758, smith1998using} and 
cryptography~\cite{yee1994using} has been explored in the context of other computing systems, we are unaware of any efforts to apply them to IWMDs. 
Here we employ the safety coprocessor as a monitoring device that observes the operations of the main processor and forewarn about and/or block unsafe operations/behaviors that can harm the user.  
To realize this, we integrate a low power coprocessor into the IWMD in such a away that it has full visibility to the activities of the main processor, 
the sensor outputs and the actuator commands. Using this information the coprocessor can sanction the actions of the main processor and take action to safeguard patient health if necessary. 


The contributions of this paper are as follows:

\begin{itemize}
\item
We propose a HW/SW framework for improving safety of IWMD operations, that consists of two main parts: a set of rules and \textit{rule-check mechanism}. The rules define the expected, safe operation of the IWMD. The rule-check mechanism checks if the medical device adhere to these rules.

\item
We propose a set of standards that developers can use to generate an effective set of rules that defines the expected, safe operation of the IWMD

\item
We propose a 2-step mechanism to test the IWMD firmware against these rules by using an existing source code analyzer and a safety coprocessor. The first step focuses on the correctness of the semantics of the program while the second step focuses on the correctness of the runtime behavior 	

\item
We design and implement a safety-enhanced IWMD controller board based on the proposed design, using a physically isolated low-power safety coprocessor.
As a case study, we implement a prototype artificial pancreas using the proposed IWMD controller board and integrate it with a physiological 
simulator running on a PC that emulates the human body model. We demonstrate the effectiveness of the proposed architecture against various practical
safety risks and evaluate its overheads.

\end{itemize}

The rest of this paper is organized as follows. We discuss of unsafe operations 
in IWMDs in Section~\ref{sec:unsafe_operations} which provides the motivational 
background for our work and introduce related previous efforts in Section~\ref{sec:realted_work}. In
Section~\ref{sec:architecture}, we propose the HW/SW architecture and describe
its mechanism of safety assurance. In Section~\ref{sec:prototyping}, we present
the design of the proposed framework and a prototype insulin pump.
Section~\ref{sec:exp_eval} describes our experimental results, and Section~\ref{sec:case_study} presents a case study of an insulin pump.
Finally, we conclude this paper in Section~\ref{sec:conclusions}.

\section{Unsafe Operation of IWMDs}\label{sec:unsafe_operations}

The major causes of unsafe operation in IWMDs are user errors, software bugs
and malicious attacks.
In insulin pump therapy, user errors contribute to a significant portion of adverse events~\cite{Zhang-JDST10}.
A study issued by
the FDA and pointed out software bugs as the major source of safety risks in medical
devices~\cite{Alemzadeh-SP13}. For example, a firmware malfunction in a BG measurement device was found to result in unexpected behaviors at BG
levels of 1,024~mg/dL or higher~\cite{Z-2544-2014}. Recent years have also seen
the emergence of security attacks on IWMDs from a remote possibility to an immediate concern. Quite recently, the National Vulnerability Database (NVD) published reports regarding two issues identified in the firmware of some cardioverter defibrillators that allows attackers to log into the IMD without the need of any passwords and disrupt its behavior \cite{nvd_2019_1}, \cite{nvd_2019_2}.
In addition, due to the stringent energy constraints and unique usage model of IWMDs, their wireless
channels are often designed without strong cryptography~\cite{Li-Healthcom11, Marin-CODASPY16}. As a result, various
vulnerabilities have been exploited to demonstrate successful security
attacks~\cite{Li-Healthcom11,Radcliffe-BlackHat11,Halperin-SSP08, Rushanan-SP14,hern_2017,newman_2018}.
Hence, security is increasingly becoming a focus of regulatory
concern~\cite{Ray-DnT15,FDA14}.

The results of both firmware malfunctions and security attacks may be catastrophic in
health-critical or life-supporting IWMDs.
For example, in the case of an insulin pump, incorrect readings from a BG meter may cause the
insulin pump to infuse lesser insulin than needed, resulting in hyperglycemia.
On the other hand, simply replaying a pre-recorded bolus dose infusion request can
lead to excessive insulin infusion, which results in hypoglycemia.
While it is easy to provide many such examples of unsafe operation,
distinguishing between safe and unsafe behavior in general is a significant
challenge. In this regard, we make a key observation that combining multiple sources of
information enables more effective and discriminatory safety checking.
For example, a bolus insulin injection command may be safe or not,
depending on the amount of insulin to be injected and the patient's current
BG level. Similarly, the control path taken within the device
firmware can be used to infer whether an insulin injection command was caused
by a command packet received over the wireless channel, or was autonomously
issued by the pump based on the sensed BG level.
Therefore, a combination of the information gathered from several sources lets us distinguish between safe and unsafe operations. As a result, a set of platform independent rules that can identify safe and unsafe operations can be generated. 
In the case where a rule has been violated by the IWMD, additional steps must be taken by the system to ensure that the patient is not affected or the effects on the patient are mitigated until proper care is taken.

\color{black}




\section{Related Work}\label{sec:realted_work}
The past few years have witnessed great interest in the reliability and security of IWMDs.
Due to the increasing complexity of IWMD firmware, empirical testing may be impractical or unable
to guarantee the integrity of the firmware~\cite{Lee-Computer06}. 
Formal methods have been proposed in this context, but the industrial adoption of this
approach is impeded by legacy hardware and software for which formal models do not exist,
and the explosion in state space that makes formal verification infeasible~\cite{Burleson-DAC12}. 
Run-time monitoring techniques that look for anomalies in program control
flow may detect malfunctions or security
attacks~\cite{Nakka-DSN04,Arora-DATE05,Mohan-HiCoNS13}. However, conventional monitoring techniques
do not take the physiological context of IWMD operations into account, and hence are not able to make precise decisions on their safety.

Various techniques have been proposed to cryptographically protect the wireless communication of IWMDs, including transferring a cryptographic key over a physically secure non-RF side channel~\cite{Halperin-SSP08, Chang-HealthSec12, Kim-DAC15} and generating one from physiological signals~\cite{Venkatasubramanian-INFOCOM08, Xu-INFOCOM11, Rostami-CCS13}.
To protect legacy IWMDs that do not utilize cryptography, or to offload power-hungry security functions from IWMDs, external devices dedicated to IWMD protection have been proposed~\cite{Gollakota-SIGCOMM11, Xu-INFOCOM11, Zhang-TBCAS13}.
These devices protect the IWMD communication channel from security attacks, but are not capable of capturing firmware malfunctions that occur internally.

Another area of research on improving the safety of IWMDs is verification through modelling. Here, a model of the IWMD is created using a modelling tool. This model is tested against several safety requirements and the transitions between different states are analyzed for 'unsafe transitions' \cite{jiang2012modeling}. The states that lead to hazardous conditions in the state space are specified as 'unsafe states' so that transitions leading into such a state can be investigated and necessary changes can be done to the firmware to avoid such transitions. Although this method identifies potential shortcomings/threats of the IWMD, it does not guarantee patients' safety from external hazards such as hijacking \cite{newman_2018}.

Usage of a watchdog unit to monitor the IWMD main microcontroller is another technique adopted to improve the security of an IWMD \cite{sieracki2008failsafe}. A watchdog unit monitors periodic signals from a programming device during programming of an IWMD. If it stops receiving
these signals, it places the IWMD in a known, safe state such as suspend the delivery of therapy or recall a known, safe delivery program. The main shortcoming of this method is that it does not safeguard the IWMD from other types of hazards. Hence, although effective, it can secure the patient only in a certain scenario.

\section{HW/SW Framework for Improving Safety of IWMDs}\label{sec:architecture}

In spite of substantial prior research efforts, there remains a gap between the need for safety assurance mechanisms in IWMDs and the state-of-the-art. Specifically, there is a need for a safety assurance mechanism that (i) is capable of capturing both device malfunctions and external attacks, (ii) makes precise decisions on safety with awareness of the physiological status, and
(iii) is easy to integrate into the HW/SW of existing IWMDs.
The design proposed here addresses all three of these concerns. The generation of rules considers all the vital information related to the control flow of the device firmware, its inputs and outputs, as well as patient data. The static rule-check can be done as an extra step during the development stage of the firmware (as explained in section \ref{sec:framework} and the safety coprocessor can be integrated into an existing device with minimal modifications to the existing device architecture.   

In this section, we first discuss the key requirements that an IWMD safety assurance framework should satisfy. These include the requirements for the rules and the requirements for the safety coprocessor.
Then, we propose a HW/SW architecture that includes the safety coprocessor and detail its operation, including the safety rule checking performed by the coprocessor to detect and prevent various unsafe operations.

\subsection{Key Requirements for the rules}\label{sec:properties_req}

To guarantee that the patient is protected from unsafe operations, the safety rules should satisfy the following requirements.

\begin{itemize}
\item \textbf{Physiological/device context:} Every rule should be defined based on one or more physiological properties and device properties. This ensures that rule checking has a positive contribution towards the patient's health and not just a superfluous process. 
\item \textbf{Inclusion of all inputs:} The outputs of the IWMD
are derived based on the inputs to the system. Therefore, the rules should consider all the inputs to the system. This allows the rules to be more robust and efficacious in checking the safety of the IWMD outputs.

\item \textbf{Inclusion of all outputs:} The commands to the actuators of the IWMD directly impact the safety of the patient. Hence it is necessary to capture all the outputs and their full operating range when defining the rules of the safety framework. This is vital, as failure to do so can have dangerous consequences on the patient's safety.  
\end{itemize}

In essence, these rules must be relationships between the inputs and outputs of the IWMD, based on physiological properties.

\subsection{Key Requirements for the rule-check mechanism}\label{sec:mechanism_req}
Although the rules may have been properly defined, if the rule-check mechanism does not perform well, then the supposed protection cannot be delivered. Therefore, we have identified that the following two requirements must be met by a productive rule-check mechanism. 

\begin{itemize}
\item \textbf{Rigorous:} The mechanism should check each and every rule meticulously. If a certain rule and its use cases are left without being checked, then the safety of the patient cannot be guaranteed, especially once that use case occurs in a real world scenario.

\item \textbf{Inconspicuous:} The rules and the procedure for checking their satisfiability should not in anyway interfere with the main operation of the IWMD, except to obstruct an unsafe IWMD operation that clearly violates one or more of the rules. Only such an operation must be prevented from being carried out. 

\end{itemize}

\color{black}

\subsection{Key Requirements for IWMD Safety coprocessor}\label{sec:principles}
To address the aforementioned safety challenges, any safety coprocessor should satisfy the following key requirements:

\begin{itemize}
\item \textbf{Visibility:} Visibility is the ability to obtain access to the necessary operational state of the IWMD in order to make inferences about safety. Safety of an IWMD operation is significantly determined by the current physiological status of the patient as well as how the device performs the operation. Therefore, to make accurate decisions, the safety mechanism should have visibility into both the extrinsic state (i.e., the patient's physiological parameters sensed by the IWMD) and the internal state of the IWMD itself (i.e., the activities of the main microcontroller).

\item \textbf{Accessibility:} Accessibility is the ability to take control of appropriate points in the IWMD immediately.
Once an unsafe operation is detected, it must be blocked before it actually acts on the patient's body.
In most cases, IWMDs inject either a drug or an electrical signal into the patient's body through the actuators.
Therefore, the safety mechanism should be able to stop unsafe actuator commands before they are executed.

\item \textbf{Isolation:} The safety assurance mechanism should be isolated from the medical functionality itself.
Separating safety functionality from complex medical functionality provides the benefits of i) minimal modification of the existing medical functionality, ii) ease of verification of the safety assurances at design time, and iii) prevention of failure propagation to the safety-critical functionality.
However, care must be taken to balance isolation with the previous two requirements because an isolated safety mechanism may have limited visibility and accessibility.
\end{itemize}

The HW/SW architecture we propose in this paper is designed and implemented  to conform with the requirements of visibility, accessibility, and isolation.
In the rest of this section, we describe the details of the design considering these requirements.


\subsection{IWMD safety framework}\label{sec:framework}

\begin{figure*}
\centering \includegraphics[width=\hsize]{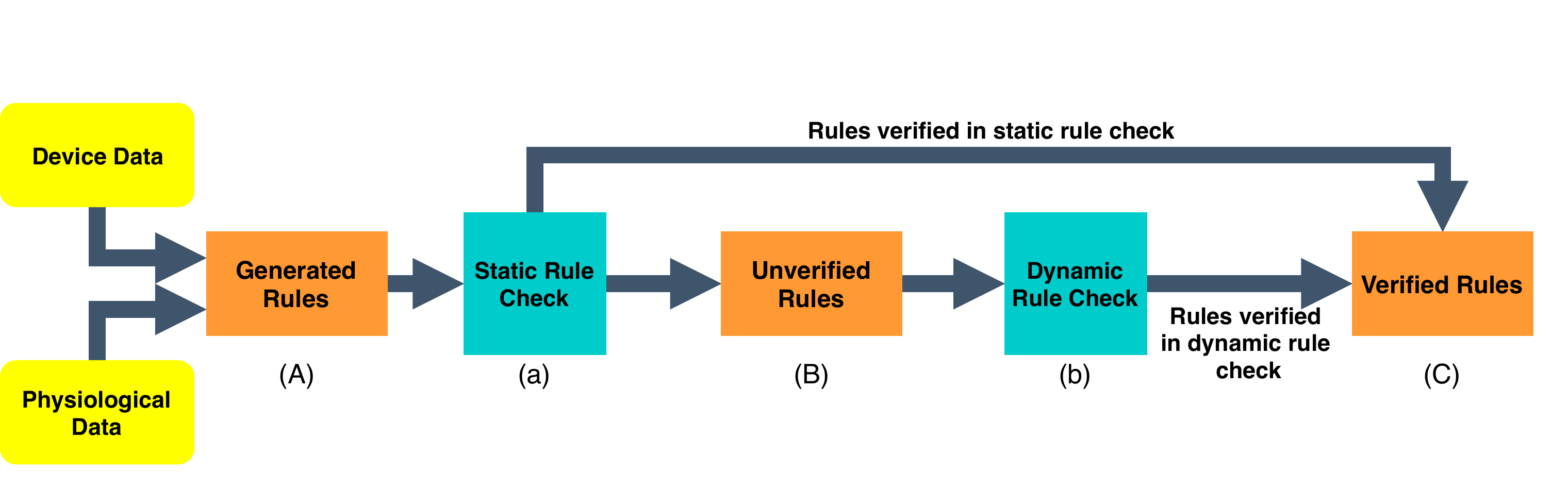}
\caption{Flow of operations of the proposed IWMD framework \label{fig:flowchart}}
\end{figure*}

Fig.~\ref{fig:flowchart} shows the steps in developing the proposed safety framework. 

The first step, step (A) in the figure, is to generate a set of rules using the physiological data and device specific data. 
This process must be done manually, preferably by someone with a sound knowledge of the physiological context and the IWMD. 
After the rules are generated, they need to be checked for validity. As mentioned in section \ref{sec:introduction}, this process has two stages. 

The first stage, static-rule check (i.e. stage \textit{a} in Fig.~\ref{fig:flowchart}) , takes place in the firmware development stage. This is done after the completion of the firmware but before deploying it on the hardware. Since the firmware is usually quite complex, doing this manually could be monotonous and impractical. Therefore, the best course of action is to use a source code analyzer or a property checker. Such a tool can do an exhaustive analysis and decide whether a certain rule is valid based on the semantics of the code. The set of rules that are validated in this step are valid regardless of the values of the sensor inputs, parameter values or the intermediate variables of the program. 
Hence, we can safely conclude that the IWMD does not break these rules anywhere in the program flow and avoid checking them in the future. As a result, these rules are removed from further validations. 

The outcome of this stage mainly depends on how well the rules are described and the robustness of the tool used. If the rules are not well described or do not capture enough vital information, then they are of little use and the analyzers may fail to validate them. Similarly, if the tool is not powerful enough, the number of rules validated is low, whereas more rules are validated if a stronger property checker or an analyzer is used. In addition, one can submit the unverified rules to a different analyzer tool so that more rules are validated in stage \textit{a}. Some commercially available source code analysis tools already employ several analyzing methods in their tools so that the repeated assertions are done automatically.  

The set of unverified rules returned from stage \textit{a} (i.e. (B) in Fig.~\ref{fig:flowchart}) must now be validated in stage \textit{b}: dynamic rule-check. This stage takes place during the actual runtime of the firmware. Everytime the main processor issues a command to the actuators of the device based on the inputs and the state of the system, that decision must be checked against the unverified rules. To accomplish this, the coprocessor is integrated into the device architecture as explained in section \ref{sec:coprocessor}. The final outcome of this procedure is a valid set of rules (step (C)) in Fig.~\ref{fig:flowchart}.

\color{black}
\subsection{Integrating the coprocessor }\label{sec:coprocessor}
\begin{figure}
\centering \includegraphics[width=10cm]{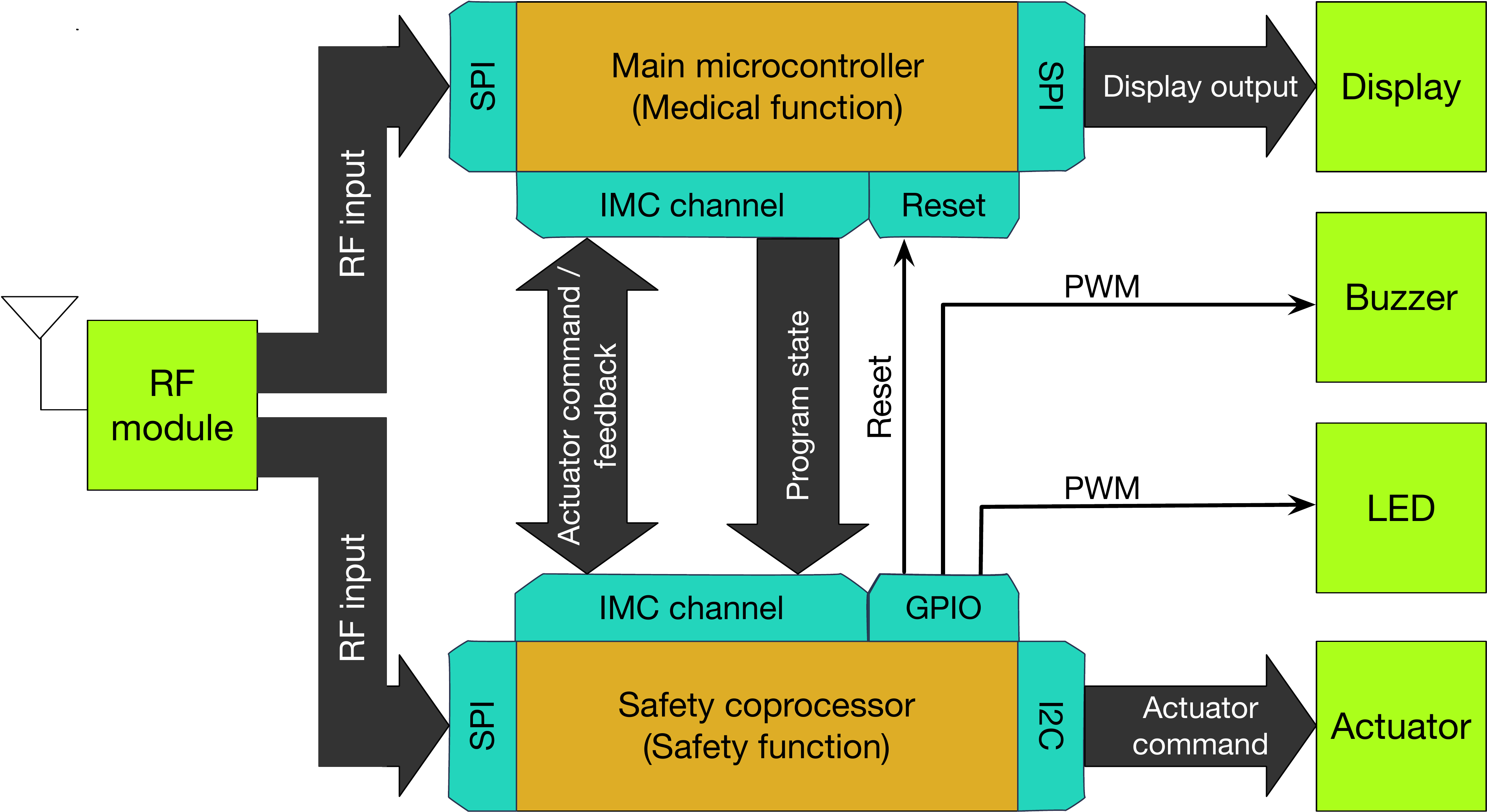}
\caption{Hardware block diagram of the proposed IWMD architecture.\label{fig:architecture}}
\end{figure}
Fig.~\ref{fig:architecture} shows the hardware architecture of the proposed safety framework. It has typical components of an IWMD, including a main microcontroller, a sensor (in a separate inter operating device that communicates over the wireless channel), and an actuator.
The medical functionality is performed by the main microcontroller, taking sensor inputs from the RF module and issuing necessary actuator commands. The main role of the safety coprocessor is to monitor various device operations, determine the safety of the operations, and block them if determined to be unsafe.
The safety coprocessor has high visibility into both the patient's physiological state as well as the device behavior.
First, it monitors the sensor input received from the RF module by snooping on the on-board communication channel.
In IWMDs where the sensor is integrated in a single device, this is done by snooping the sensor-to-main microcontroller communication channel.
It only listens to the communication in a passive manner and does not interfere unless an unsafe operation is detected.
At the same time, it also monitors the program state of the firmware running on the main microcontroller over a dedicated communication channel called the inter microcontroller communication (IMC) channel.
The main microcontroller generates messages whenever the program state changes (e.g., entering and exiting of critical functions) so that the safety processor can keep track of it.

Unlike that the sensor inputs are only passively monitored, actuator commands are intercepted in the middle of the main microcontroller and the actuator by the safety coprocessor.
The intercepted actuator commands are first inspected by the safety coprocessor and passed on only when their safety is confirmed.
If the safety coprocessor determines the command is not safe, it blocks the command and performs necessary actions as described in Section~\ref{sec:follow-up}.
The rerouting of actuator commands grants the safety coprocessor with ability to counteract potential unsafe operation in a timely manner, before it takes actual effect.
While the main controller no longer has direct control over the actuator, this rerouting is transparent to the main microcontroller, and thus it requires only minimal hardware modification.

In this architecture, even though the safety coprocessor is physically isolated from the main microcontroller and does not actively share any recourses, it has high visibility and accessibility necessary to detect and prevent potential unsafe operations.
Hardware modification of this architecture is minimal since the safety coprocessor is the only additional hardware and rerouting or tapping off-chip communication buses does not incur significant overheads.
Software modifications required in the main microcontroller include generating a unique message when the program state changes and removing/changing the most recent operation in the records in the case it is deemed unsafe and interrupted by the coprocessor. These are minor modifications with negligible influences on its execution. 
In the next section, we describe how this framework used to applied to specific IWMD using a reference IWMD system. 

\begin{figure}
\centering \includegraphics[width=7cm]{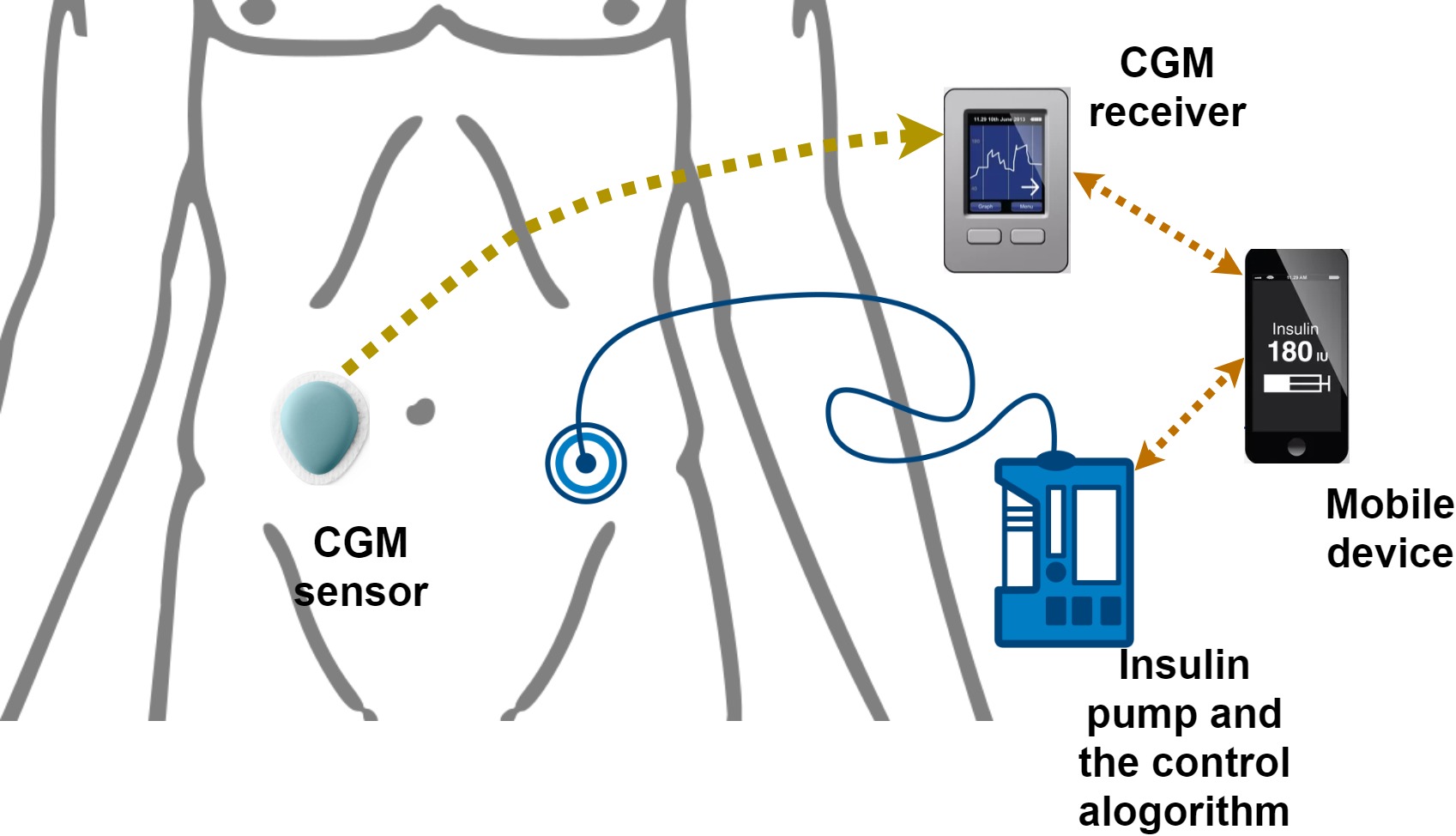}
\caption{Diagram explaining the operation of an artificial pancreas.\label{fig:bodyiwmd}}
\end{figure}

\begin{figure}

\centering \includegraphics[width=\hsize]{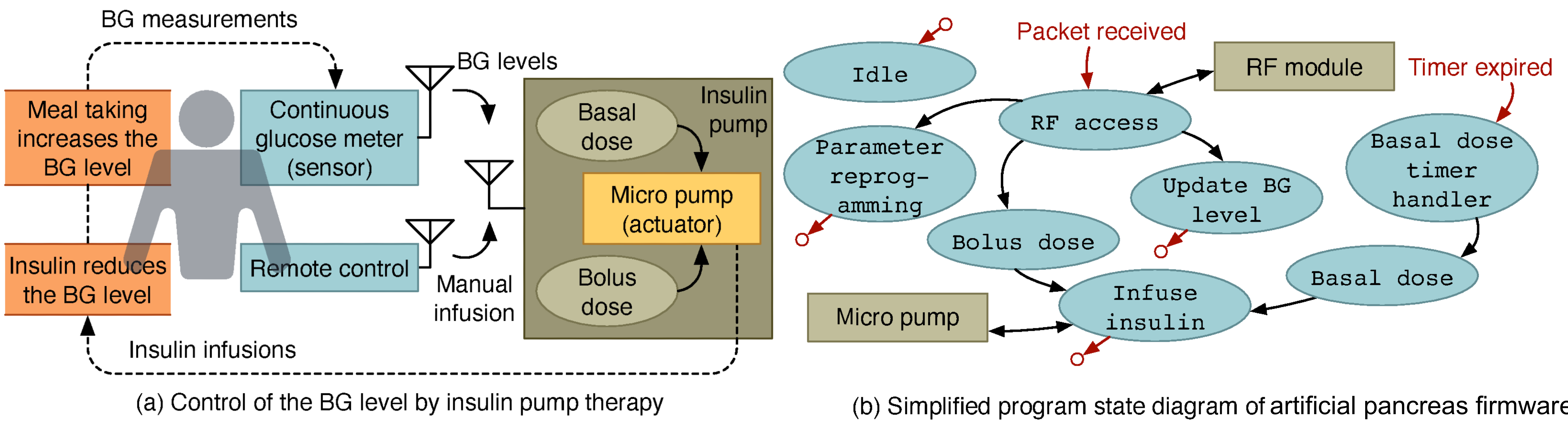}
\caption{Artificial pancreas system.\label{fig:insulin_pump}}
\end{figure}


\subsection{Design of a reference IWMD model based on the framework}

\subsubsection{Reference IWMD model}

In this paper, we consider an artificial pancreas as our reference IWMD model, which is shown in Fig.~\ref{fig:bodyiwmd}.
The four parts in the figure (including the mobile device) should work hand in hand to guarantee proper operation. 
The insulin pump (including the control algorithm) is connected via a wireless medium to a continuous glucose meter (CGM) receiver for autonomous BG level monitoring. The CGM sensor reads the BG level and sends it to the receiver.
The patient can also send a set of specific commands (such as issuing a bolus dose) using the mobile device.
The control algorithm takes all these data into consideration and calculates a suitable amount of insulin to be infused. It then commands the pump mechanism to infuse insulin if required.
Hence, this is a good platform to test our safety framework because such high interoperability and frequent user interactions raise reliability and security concerns in these devices.
Although we take an artificial pancreas as a reference IWMD, the proposed safety framework is also applicable to any type of IWMD.
Next we describe briefly the operation of an artificial pancreas.
\color{black}
\subsubsection{Basics of artificial pancreas operation}\label{sec:insulin_pump_basics}
People with type 1 diabetes, who have a problem with controlling the BG levels due to the pancreas not producing enough insulin, often rely on an insulin pump to continuously deliver artificial insulin through a catheter under the skin.
The control of BG levels by an artificial pancreas system is illustrated in Fig.~\ref{fig:insulin_pump}(a).
The CGM sensor periodically measures the BG level and wirelessly transmits the measurements to the CGM receiver and the receiver transmits this to the control algorithm.
These data are used to automatically calculate the basal dose rate (insulin needed to keep BG levels constant during periods of fasting) and to help the patient calculate bolus doses (insulin need to reduce BG levels increased by taking a meal).
About 15  to 30 minutes before each meal the patient manually requests an infusion of a bolus dose calculated as follows to cover the estimated carbohydrate in the meal and to prevent sudden elevations of the BG level:
\begin{align}
\textit{(bolus dose)} =~&\textit{(carbohydrate coverage dose)}  \nonumber \\
& + \textit{(correction dose)} - \textit{(active insulin)} \nonumber
\end{align}
The carbohydrate coverage dose is to cover the additional carbohydrate taken from the meal:
$$\textit{(carbohydrate coverage dose)} = \frac{\textit{(grams of carbohydrate)}}{\textit{(carbohydrate ratio)}}$$
The correction dose is to adjust current BG level to the target BG level:
$$\textit{(correction dose)} = \frac{\textit{(current BG) - (target BG)}}{\textit{(correction factor)}}$$
Finally, the active insulin (or insulin-on-board) is the amount of insulin that is still active in the body from the previous bolus doses.

The algorithm for calculating the basal dose in our reference device is based on \cite{openaps.org}. 
In brief, this algorithm takes the current and previous CGM readings and the previous doses of insulin (both basal and bolus) into account, along with data of any meals taken recently. Using these data, it generates an accurate assessment of the behavior of the blood glucose, which in turn is used to calculate the amount of insulin required.

In the next section we look into how our proposed framework can be integrated into the architecture of the artificial pancreas.

\subsection{Applying the framework to the reference IWMD model }\label{sec:applicationOfframeowrk}

\subsubsection{Generating rules}\label{sec:coprocessor_rule_checking}
 This is the first step in our proposed framework shown in Fig.~\ref{fig:flowchart}. Specifying robust safety rules based on the expected, safe device behavior is the first and most critical step of the safety assurance. 
 For the purpose of this paper, we are considering an artificial pancreas system that is being used for treating a patient with type I diabetes. 
 Table \ref{tab:genRuleData} shows the types of data that is taken into account along with some examples related to our reference model. Here, the physiological data represents the diabetes related data. The state variables are a type of device data that describes the state of the program. The Fig.~\ref{fig:insulin_pump} manifests the simplified program state diagram of the firmware of an artificial pancreas. The ovals denote the states of the program execution and the squares denote the I/O components sued by the program. The transitions between states or I/O components accessed from each state are represented by the edges. The inputs to the device are collected from two sources: the CGM sensor and the wireless access point that receives commands from the user. Finally, the outputs of the system are the commands issued by the firmware to the device's actuators.
 
\begin{table}
  \caption{Data for generating the rules}
  \label{tab:genRuleData}
  \begin{tabular}{l|l}
    \toprule
    Type of data & Example\\
    \midrule
    Physiological data & Safe BG range, unusual BG levels, maximum \\ 
                       & dose of basal, bolus insulins, minimum time \\
                       & interval between two consecutive insulin doses \\ 
    \hline
    Inputs to the device &  BG level, bolus dose, meal intake\\
    \hline
    State variables of the device & Shown in Fig.~\ref{fig:insulin_pump} \\
    \hline
    Outputs to the device & Commands to the insulin pump, set of LEDs,\\ 
     &  buzzer and optional LCD display \\
   \bottomrule
\end{tabular}
\end{table}

Once the appropriate information has been identified and with the help of physiological knowledge and device specific information, the next step is to define the safety rules based on the relationships between these parameters/variables. The rules are defined under four categories as follows:


\begin{figure}
\centering\includegraphics[width=8cm]{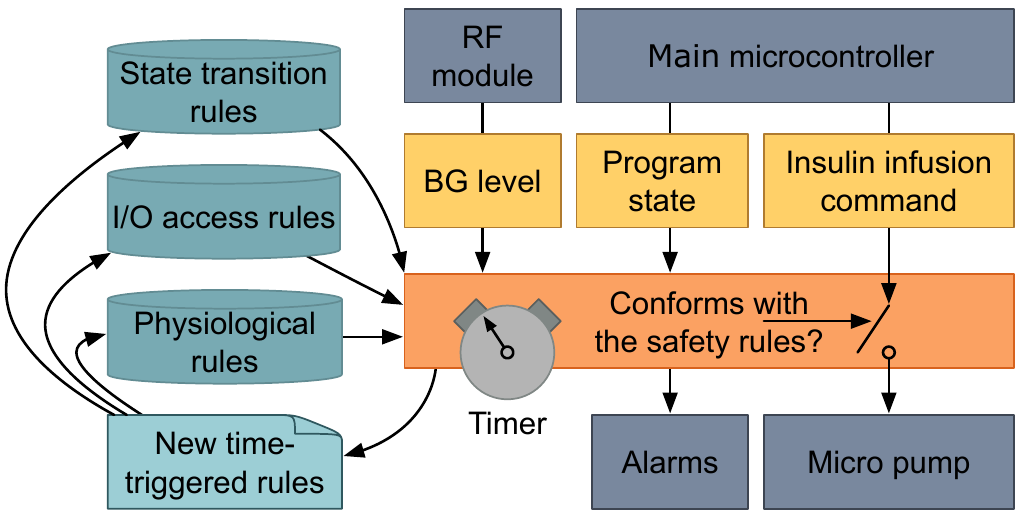}
\caption{Safety rule checking performed by the safety coprocessor.\label{fig:safety_rule_checking}}
\end{figure}

\color{black}
\begin{itemize}

\item \textbf{I/O access rules:}
An unexpected access to I/O components suggests the potential existence of firmware malfunctions security attacks.
I/O access rules define legal I/O accesses to peripheral components, such as the sensor, actuator, RF module, off-chip memory, etc., for each state.
For example, the micro pump access is legal only when the main microcontroller is in the \textit{Infuse insulin} state, only once per entrance into this state.
The access pattern to the I/O components is also to be examined by the rules.
For example, ceaseless RF access attempts intended to prematurely drain the battery can be detected by limiting the RF module access frequency.
I/O access rules do not define the range of actuator operations, i.e., the amount of insulin infusion, which is to be checked by the physiological rules.

\item \textbf{Physiological rules:}
As discussed in Section~\ref{sec:insulin_pump_basics}, and as shown in Fig.~\ref{fig:insulin_pump}(b) as well, an insulin infusion can be triggered by multiple modes: basal insulin or bolus insulin.
Depending on what mode has triggered the insulin infusion, the safe range of insulin amount is different.
The program state reported by the main microcontroller, which is first checked by the state transition rules, is used to infer the mode of operation.
Then, the BG levels obtained by monitoring the RF module activities are utilized to see if the insulin infusion command issued by the main microcontroller is safe or not, considering the inferred mode of operation.

\item \textbf{State transition rules:}
State transition rules define the frequency of entering each state, time from entering to exiting each state, and legal next states for each state.
Checking these rules can detect firmware malfunctions or security attacks that change the execution paths of the program.

\item\textbf{Time-triggered rules:}
Some rule enforcements are performed not right after the occurrence of a device activity, but after some time has elapsed from it.
A time-triggered rule is dynamically generated in any of the three types of rules above.
If an expected state transition or I/O access is not observed within a predefined time frame, this implies that the main microcontroller or peripheral components are not responding, which can be captured by time-triggered state transition rules or I/O access rules.
Time-triggered physiological rules can be used to monitor changes in the physiological status periodically or after the actuator is operated.
In addition, this type of a rule can also be used to check the accessibility to the main microcontroller, as explained in section \ref{sec:rule_check_mech}.

\end{itemize} 

\begin{figure}
\centering \includegraphics[width=\hsize]{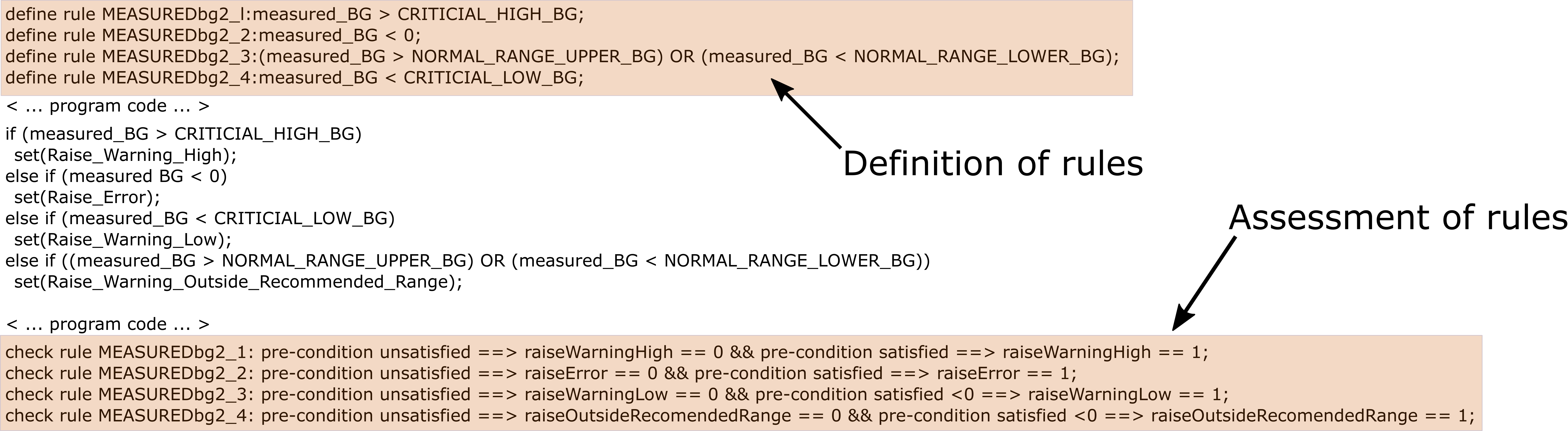}
\caption{Snippet of pseudo code from static rule-check checking.\label{fig:static_rule_check}}
\end{figure}

\begin{figure}
\centering \includegraphics[width=10cm]{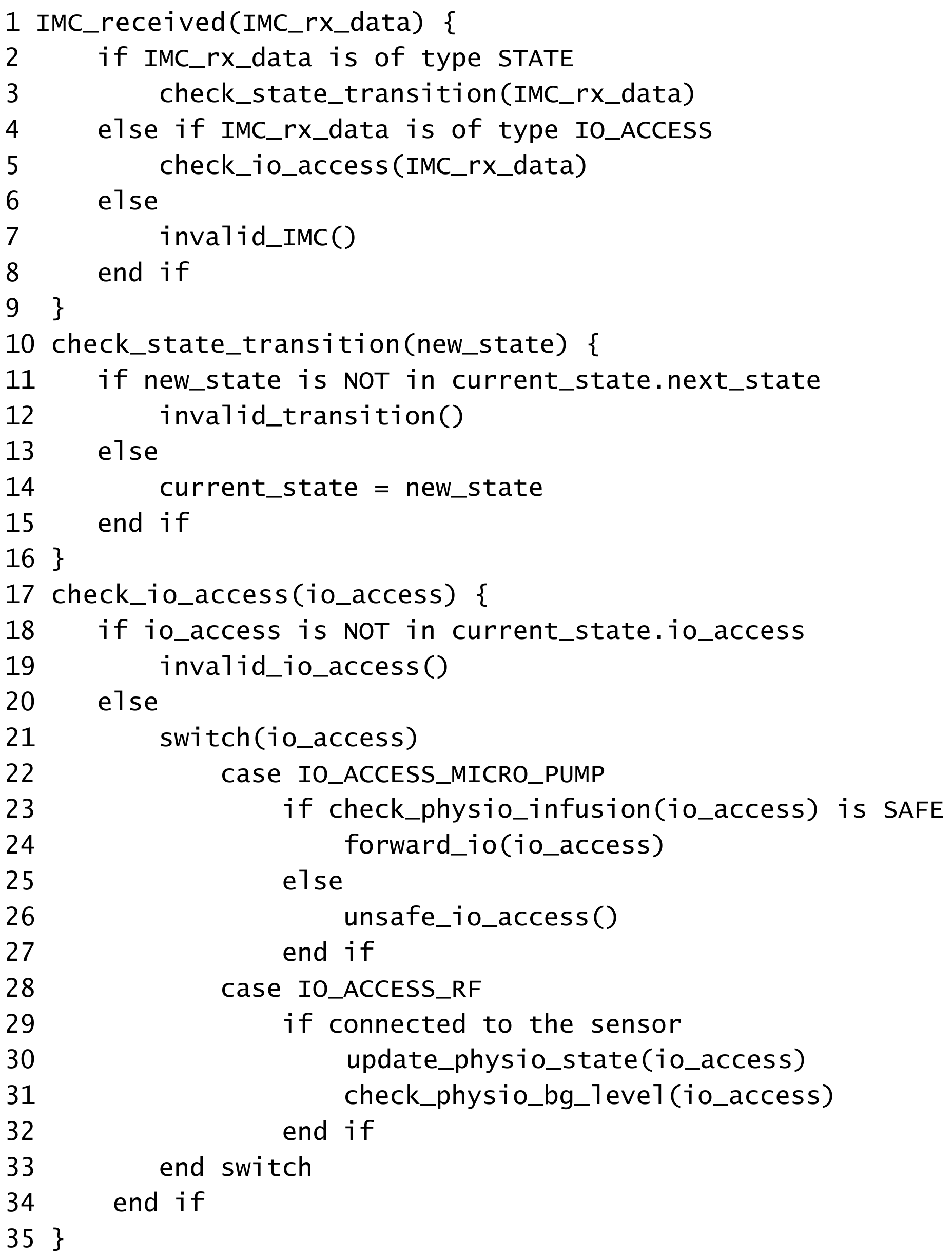}
\caption{Pseudo code of the state transition rule checking and the I/O access rule checking.\label{fig:code_snippet_1}}
\end{figure}


\subsubsection{Rule-check mechanism}\label{sec:rule_check_mech}
After generating the rules and completing the source code for the firmware, the next step is to check the satisfiability of the rules. As mentioned in section \ref{sec:framework}, this consists of two steps. 
The first step, static-rule checking (stage \textit{(a)} in Fig.~\ref{fig:flowchart}), is done in the firmware development stage using an existing, off the shelf property checker or a source code analyzer. For the purpose of this project, we are using Frama-C framework \cite{cuoq2012frama} that is specifically designed for software written in C programming language. 
Fig.~\ref{fig:static_rule_check} shows a snippet of the pseudo code that contains definitions of 4 rules and the respective lines of the code. The first highlighted part defines the rules. The sention below that is the actual code and highlighted section below that checks if the above rules can be validated or not using the given pience of code. Note that the rules shown here are fairly simple, but are very crucial for guaranteeing the patient's safety. 
The assertions in the given pseudo code pertains to the following rules: 

\begin{itemize}
\item Raise the error message if the measured BG level is below zero
\item Raise the warning 'BG LEVEL VERY HIGH' if the BG level is above the critically high BG level
\item Raise the warning 'BG LEVEL VERY LOW' if the BG level is below the critically low BG level
\end{itemize}
In the source code, the rules need to be added in the format required by the analyzer tool.
All the rules that were generated should be added where appropriate and tested. 

Once the static-rule check is completed, some of rules would be validated while some are not. The rules that were validated, called 'statically valid rules', are no longer a concern as it has been proven that regardless of the variables values during run time, the firmware will definitely operate in a safe manner as far as the statically valid rules are concerned. 
However, rules that were not asserted in the above step, usually the ones that require complex calculations, real time data or program state cannot be asserted in a similar manner. Hence, these rules have to be asserted during run time using the safety coprocessor.

\begin{figure}
\centering \includegraphics[width=8cm]{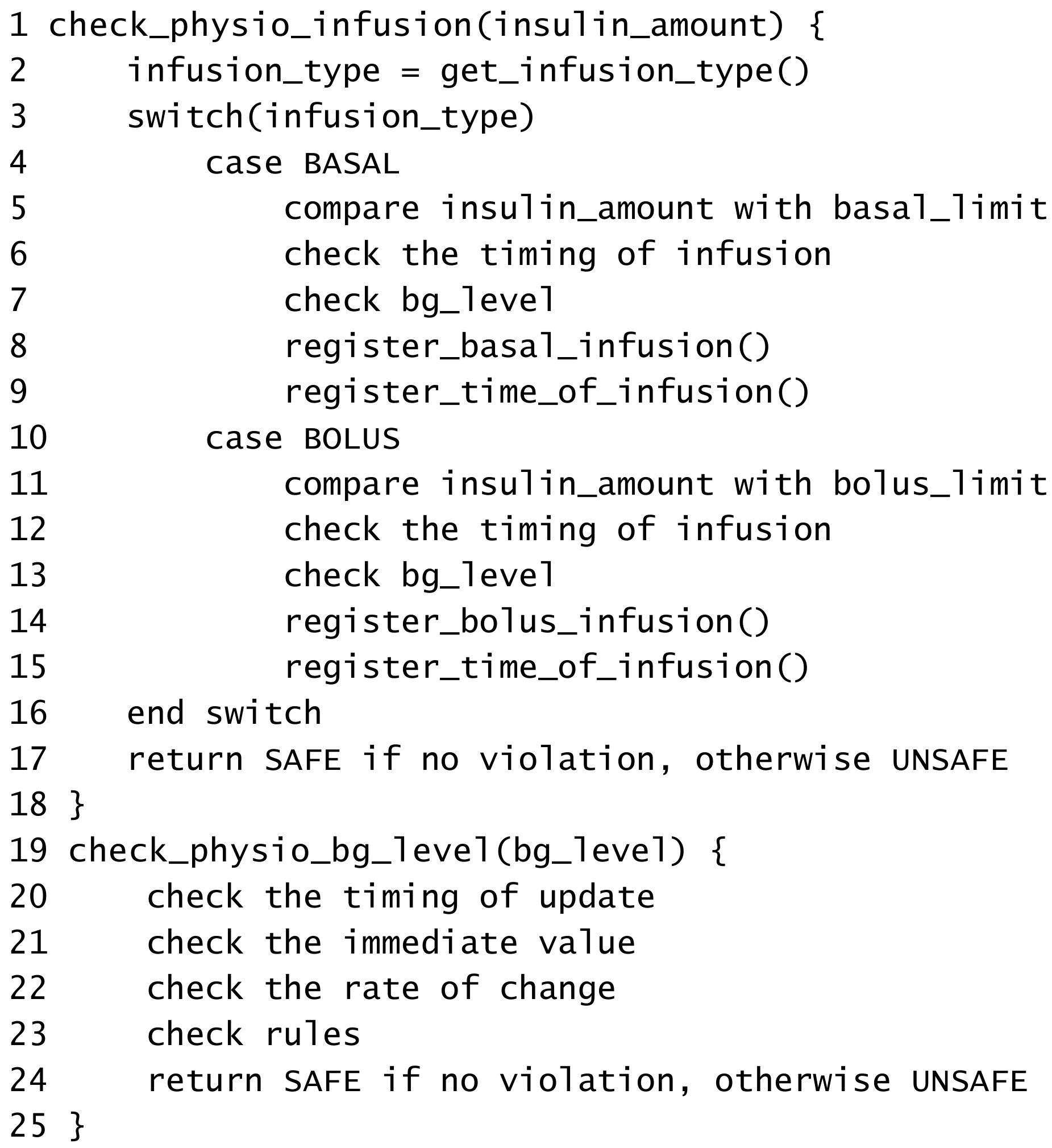}
\caption{Pseudocode of the physiological rule checking.\label{fig:code_snippet_2}}
\end{figure}

\color{black}
Compared to the main microcontroller's firmware that performs complex medical functionality, the safety rule checking program deployed on the safety coprocessor is relatively simple.
Fig.~\ref{fig:code_snippet_1} shows the pseudo code of the state transition rule checking and I/O access rule checking running on the safety coprocessor.
First, when a message \texttt{IMC\_rx\_data} is received from the main microcontroller via the IMC channel, the type of the received message is first determined by \texttt{IMC\_received()} (Line 1).
For a state transition report message, the \texttt{check\_state\_transition()} function updates the current state if it is a legal transition (Line 12--16). 
It should be noted that although a static rule check can validate proper state transitions, it cannot guarantee that an illegal state transition due to an external influence does not occur \cite{vishwakarma2018exploiting}. Hence to guarantee safety, we decided to include a simple state transition check within the coprocessor.   
\color{black}
The \texttt{check\_io\_access()} function is invoked by an explicit I/O activity, such as an actuator command from the main microcontroller, or an implicit I/O access snooped from the communication between the main microcontroller and I/O devices, such as RF communication.
The function first checks if the current state has the access permission (Line 20).
Depending on the type of the I/O access, it checks the physiological safety and forwards the command to the actuator (Line 25--26), or update the physiological state obtained from the sensor (Line 32).
Fig.~\ref{fig:code_snippet_2} shows the pseudocode for checking some physiological safety rules. 

The \texttt{check\_physio\_infusion()} function is invoked by \texttt{check\_io\_access()} before actuator commands are forwarded.
From recent state transition history, the function first determines the context of the insulin infusion, and different safety rules are enforced depending weather it is a bolus or basal (Lines 4--15).
The \texttt{check\_physio\_bg\_level()} function is also invoked by \texttt{check\_io\_access()} whenever the BG level is updated from the sensor.
During the update, the value, the rate of change, etc. of the BG level is checked. 
Similarly, pseudocode for all the physiological rules considered must be generated.


The safety coprocessor can also act as a 'watchdog' to monitor the main microcontroller. It is supposed to receive a periodic message from the main microcontroller through the IMC channel. If the designated time elapses without receiving a message, then the safety coprocessor concludes that the main processor has entered a non-responsive state and takes necessary action to resolve the issue.  
\color{black}


%

\subsection{Follow-up Actions}\label{sec:follow-up}
Upon the detection of unsafe operations, the safety coprocessor performs necessary follow-up actions.
For the violations of state transition rules and I/O access rules, the safety coprocessor may raise a user-perceptible warning so that the patient can take necessary actions, such as device checkup.
Visual, audible, or vibratory alerts can be used.
It also can utilize an external device (e.g., smartphone) to provide more details about the problem.
If the violations occur persistently and may damage the device (e.g., battery drain attack), the safety coprocessor may temporarily disable and isolate the problematic components.
In a situation where the main processor stops responding (e.g., hangs) the safety coprocessor can restart the main microcontroller by asserting the reset signal.
In addition, the safety coprocessor can be designed to perform minimal medical functionality to give time for the patient to take necessary actions.
For example, in an artificial pancreas, a pre-programmed basal insulin can be infused at a fixed rate, while bolus doses are manually infused using an emergency insulin pen until the problem is resolved by the medical professional. However, we believe that expert knowledge of the context is required to design effective countermeasures. Hence in this paper we have limited our scope to identifying, blocking and informing the user about unsafe behavior. Extending the proposed design to include countermeasures is trivial and requires minimal or no change at all to the current design. 

\color{black}

\section{Implementation and integration of the safety coprocessor}\label{sec:prototyping}

In this section, we first present the design of the safety-enhanced controller board based on the proposed architecture.
Next, we present a prototype insulin pump system that incorporates the controller board and integrates the system with a human body model simulator for the evaluation of safety rule checking.

\subsection{Safety-Enhanced Controller Board Design}\label{sec:proto_board}

\begin{figure}
	\centering \includegraphics[width=\hsize]{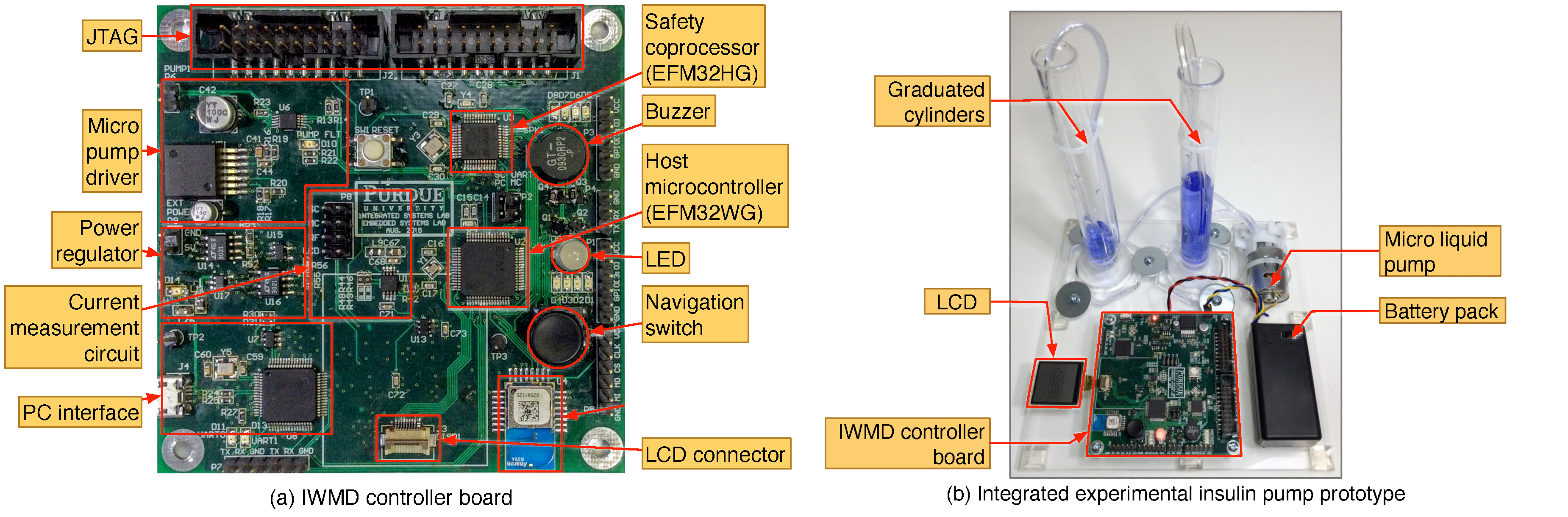}
	\caption{Photographs of the IWMD controller board and the prototype insulin pump.\label{fig:exp_photo}}
\end{figure}

Fig.~\ref{fig:exp_photo}(a) shows a photograph of the safety-enhanced controller  board.
We use heterogeneous microcontrollers for the main microcontroller and the safety coprocessor.
For the main microcontroller that performs high-complexity medical functionality, we use EFM32WG from Silicon Labs.
This microcontroller incorporates the ARM Cortex M4 core with the maximum frequency of 48~MHz, and its power consumption in active mode is 225~$\upmu$A/MHz.
The safety coprocessor is selected with an emphasis on low sleep-mode power consumption because the safety coprocessor is in the sleep mode for most of the time, waiting for an event to trigger the safety rule checks.
We use EFM32HG from Silicon Labs, based on the ARM Cortex M0+ core, which consumes only 0.9~$\upmu$A in deep sleep mode with the real-time clock (RTC) on.
When in active mode, it consumes 114~$\upmu$A/MHz at up to 25~MHz.
A Low Energy Universal Asynchronous Receiver/Transmitter (LEUART) channel is used as the IMC channel between the two micro-controllers.
It consumes very low power and only requires the RTC to operate.
The baud rate is limited to 9600 bps, but this is sufficient for our system as we will see in Section~\ref{sec:exp_eval}.

The board also has peripheral components necessary to control an insulin pump.
A 2.4-GHz Bluetooth module is used to communicate with the CGM and the remote control.
\color{black}
It is connected to the main microcontroller through a Serial Peripheral Interface (SPI) bus.
We do not employ a power-consuming crossover switch to steer the bus signals because once the safety coprocessor becomes the master device after an unsafe operation is detected, the main microcontroller is to remain inactive until the problem is fixed.
The micro pump driver is connected to the safety coprocessor through the Inter-Integrated Circuit (I$^2$C) bus.
As described in Section~\ref{sec:coprocessor}, the main microcontroller does not have direct access to the pump driver.
The buzzer and the LED are used for safety alerts, and the navigation switch and the LCD are used as user interfaces.
The PC interface is used to integrate the controller board into a PC simulation environment, as described in Section~\ref{sec:platform}.
\color{black}
The board is powered either by the USB bus power when connected to a PC or by an external battery pack.

\subsection{Prototype of artificial pancreas}\label{sec:platform}
Fig.~\ref{fig:exp_photo}(b) shows a prototype insulin pump system that incorporates the implemented safety-enhanced controller board.
An LCD, a battery pack, a set of LEDs, a buzzer and a micro liquid pump are integrated with the board to compose a complete insulin pump system.
The micro liquid pump moves the liquid from one graduated cylinder to another.
Note that this pump is not a precision pump used in actual insulin pumps but is only for demonstration purpose to visualize the infusions of insulin using colored liquid.
Hence, we do not measure the current consumption of this pump.


On the host PC, we implement a Matlab software that simulates the physiological behavior of a diabetic patient.
A well-known physiological model called Glucose-Insulin Model (GIM)~\cite{Man-JDST07} simulates the variations of the BG levels for the glucose and insulin presented.
The GIM is modified to interact with the controller board to update the BG levels and insulin infusions on the fly.
The prototype insulin pump receives BG levels and insulin infusion requests from the simulator and performs necessary insulin infusion operations.
The amount of insulin infused is notified to the simulator, and in turn, to the GIM, which updates the BG level.
In the meantime, the adversary on the host PC can generate counterfeit insulin infusion requests.
The safety rules are implemented in the simulator as well, in order to verify the functionalities of the safety rule checking performed by the safety coprocessor.

\section{Performance and Overheads Evaluation}\label{sec:exp_eval}
In this section, we measure the performance and the overheads of the proposed architecture, focusing on the impact of the introduction of the safety coprocessor.



\begin{figure}
	\centering \includegraphics[width=12cm]{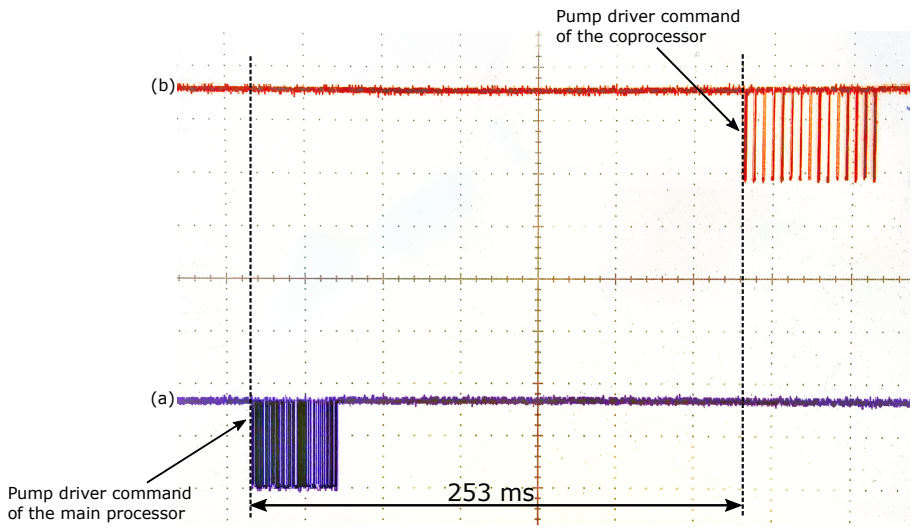}
	\caption{Waveform of (a) UART from the main microcontroller to the safety coprocessor, and (b) Command to the pump driver from safety coprocessor.\label{fig:exp_waveform}}
\end{figure}

\subsection{Impact on Performance}
Since the modifications of the main microcontroller firmware is mostly associated with IMC message generation, the performance degradation is negligible.
However, by introducing the safety coprocessor, a propagation delay is introduced to the duration actuator commands issued by the main microcontroller takes to reach the actuator.
This delay consists of the time taken to receive the actuator command by the safety coprocessor, check the safety conformance, and re-issue the command to the actuator.
Fig.~\ref{fig:exp_waveform} shows the communication waveform from between the coprocessor and main processor.
As shown in Waveform (a), the main processor sends a command destined for the pump driver.
The safety of this command is checked and reissued by the coprocessor, as shown in Waveform (b). An additional time of 253~ms is taken by the coprocessor to carryout this check. The majority of this time is spent on transmitting the message via the IMC channel.
A delay of 253~ms is negligible as insulin infusions take up to several seconds.
For IWMDs which this time delay is not tolerable, a faster IMC channel can be used to reduce the delay.

\subsection{Impact on energy consumption}
\color{black}
For an artificial pancreas running for one month with a 2500~mAh AA-sized battery, the safety coprocessor incurs 3\% of energy overhead.
Generally, the safety rule checking is computationally much simpler than the medical functionality itself, hence a lower power processor can be used.
Moreover, the safety coprocessor can be more heavily duty-cycled than the main microcontroller because the safety rule checking is triggered mainly by the main microcontroller's activities.
As a result, the power consumption of the safety coprocessor scales with that of the main microcontroller in other types of IMWDs.
In IWMDs like artificial pancreas, where the majority of the energy is consumed by  peripheral components, such as a sensor, an actuator, and a communication channel, the microcontrollers' contribution is not significant.





In addition to the above, the static-rule check step adds an unavoidable overhead to the development stage as a certain duration of time should be spent on generating, refining and asserting the safety rules. However, because it reduces the load on the safety coprocessor and contributes to the quality of the firmware and the overall product, we consider it as a reasonable trade off.

\color{black}

\section{Evaluation of the proposed architecture }
In this section, we discuss the performance of the proposed framework in several practical scenarios. First, in section \ref{sec:case_study_static} we briefly look at some results obtained during the static rule-check stage. Then from section \ref{sec:scenario1} to \ref{sec:scenario5} we look at how the proposed dynamic rule-check mechanism behaves in several practical situations that can be harmful to the patient. 

\subsection{Case Study of a static rule check}\label{sec:case_study_static}
\begin{figure}
\centering \includegraphics[width=12cm]{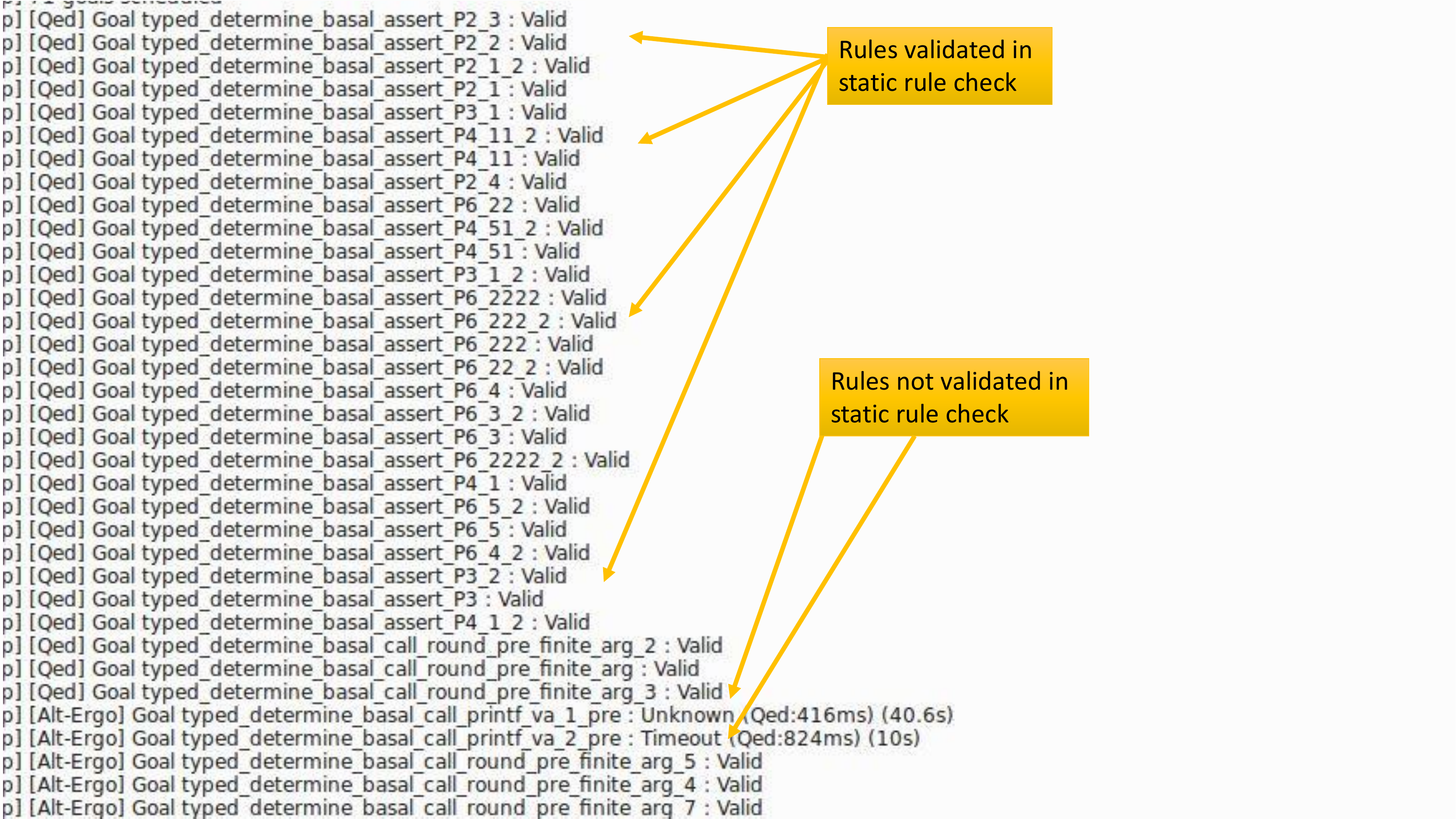}
\caption{Frama-C output of a static rule check.\label{fig:staticRulesOtPt}}
\end{figure}

Fig. \ref{fig:staticRulesOtPt} is a section of the output of the static rule check. As shown in section \ref{sec:rule_check_mech}, after the assertions are added as annotations, the source code is tested against several automated theorem provers. In Frama-C, several such provers are employed. The assertions shown in Fig.~\ref{fig:staticRulesOtPt} have been validated using Alt-Ergo 
\cite{sas_2016} and Qed \cite{mitchell_2015} provers. 
If at least one of the provers proves that a rule is valid, then it is labelled as a valid rule and is removed from further checks. We can see from the output that most rules have already been validated. Hence after concluding this test, we can safely assume that at the minimum the firmware will behave according to the validated rules. In fact, the rules mentioned in section~\ref{sec:rule_check_mech} were validated in this step. (i.e. regardless of the program behavior, the user will be warned as expected in the event the BG level is outside the safe range). Hence the next step is to try to validate the unsatisfied rules that are labelled as 'unknown' or 'timeout' in Fig.~\ref{fig:staticRulesOtPt}. These rules must be validated in the dynamic rule check stage.

\color{black}

\subsection{Case Study of dynamic rule checks}\label{sec:case_study}
In this section, we demonstrate the effectiveness of the proposed dynamic rule check mechanism through experiments using the prototype artificial pancreas introduced in Section~\ref{sec:platform}.

\subsubsection{Patient Model}
We assume a patient with type 1 diabetes in our experiments.
The default parameters of the patient defined in the GIM~\cite{Man-JDST07} are used, which are summarized in Table~\ref{tab:gim_parameters}.


\subsubsection{Safety Rules}\label{sec:defining_safety_rules}
We define safety rules as follows.
\begin{itemize}
\item \textbf{I/O access rules:}
The I/O access rules are also defined based on the the state transition diagram in Fig.~\ref{fig:insulin_pump}(b).
We monitor two peripheral components: the RF module and the micro pump.
The RF module can be accessed only from the \texttt{RF access} state, and the micro pump can be access only from the \texttt{Insulin infusion} state.
If any peripheral component is accessed from other than these legal states, it is a violation of the I/O access rules.

\item \textbf{State transition rules:}
The state transition rules are defined based on the same state transition diagram.
State transitions are legal only as denoted by the directed edges; all other transitions are identified as a violation.
We profile the execution time and define the time-triggered rules with the transition time constraint.

\item  \textbf{Physiological and time triggered rules:}
We define several physiological rules that are associated with the operation of the micro pump.
The details of these physiological rules and the unsafe operations prevented by them are described in sections ~\ref{sec:scenario1}, \ref{sec:scenario2}, \ref{sec:scenario3}, \ref{sec:scenario4} and \ref{sec:scenario5}.
\end{itemize}

\begin{table}
  \caption{Simulation parameters of a type 1 diabetes patient.}
  \label{tab:gim_parameters}
  \begin{tabular}{ccl}
    \toprule
    Parameter&Value\\
    \midrule
    Body weight &  78~kg\\
    Endogenous glucose production &  2.40~mg/kg/min\\
   \bottomrule
\end{tabular}
\end{table}


\subsubsection{Scenario 1: Recurring bolus dose requests in quick successions}\label{sec:scenario1}

The first attack scenario involves requesting bolus doses from the main processor repeatedly in a short time span. The request is sent using the mobile device via the wireless interface.

In this experiment, we exploit the bolus dose request feature through the wireless channel.
When a user is preparing to consume some food, he/she should inform the device the quantity of carbohydrate consumed and the dosage of bolus insulin that needs to administered. This is usually done about 10-20 minutes before consuming food.

As shown in the state diagram in Fig.~\ref{fig:insulin_pump} when the main processor receives a data packet over the wireless channel, it processes the packet immediately and infuses the requested the amount of bolus insulin. An attacker can exploit this weakness. Hence, when a sequence of such packets are sent, there exists a danger of causing hypoglycemia and harm the patient. In addition, an attacker can send a large number of packets to cripple the main processor and disrupt the device operations completely. A scenario of this nature is addressed in a later case study.  

\begin{figure}
\centering \includegraphics[width=8cm]{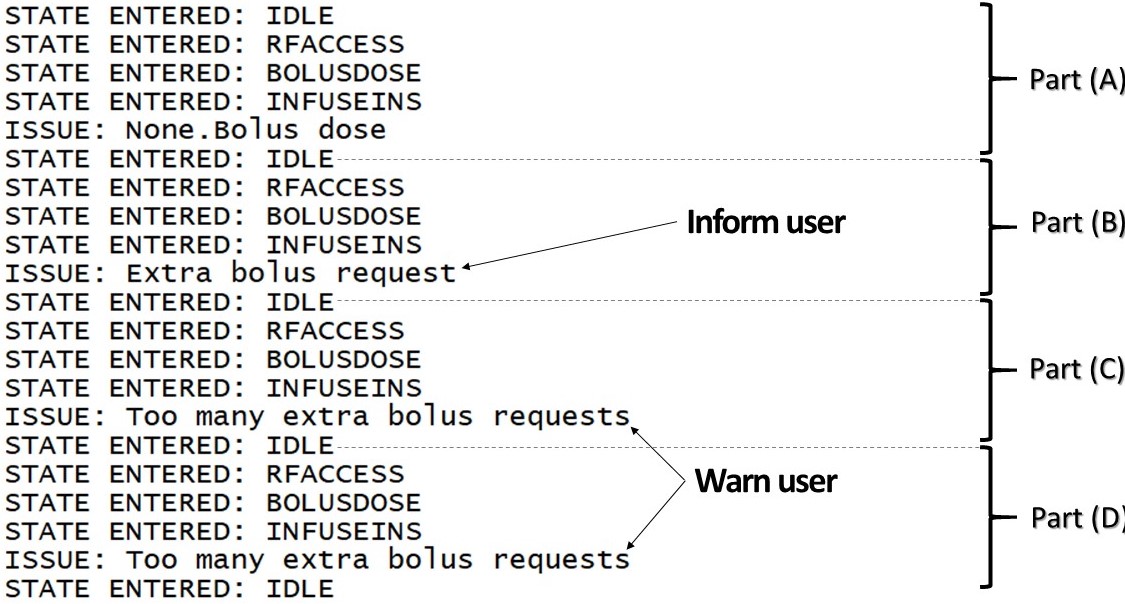}
\caption{Decoded log pertaining to the recurring bolus dose requests.\label{fig:blAttLog}}
\end{figure}

An attack of this nature can be identified by a time-triggered rule. Fig.~\ref{fig:blAttLog} is a snippet of the decoded messages of the logs generated by the coprocessor. Here, we have four back to back to bolus requests generated within a period of 10 minutes. Regardless of this, the safety coprocessor functions in a similar manner even when the bolus requests are spread in a longer duration and when there are basal doses in-between. We also assume that the user has not administered a bolus insulin dose(i.e. has not consumed any carbohydrates) during the past 3 hours.  

As shown in part (A) of the snippet, when the main processor sends the first pump message to the coprocessor, it processes the message and commands the pump to administer the insulin. The first four messages shows how the program state has changed. The message to the coprocessor is sent when the main processor is in the \textit{Infuse Insulin} state. The fifth message which says that the bolus dose was administered as requested without any issues, is generated by the safety coprocessor, when it commands the pump to administer the bolus dose.  

However, the latter requests are not processed in a similar manner. When the main processor receives the second request, it processes it in a similar manner, as shown in the state transitions in Part (B) of Fig.~\ref{fig:blAttLog}. However, when the coprocessor receives the message from the main processor, it checks previous bolus requests and decides that this is a violation of a physiological rule as the period between the two consecutive bolus infusions is shorter than the safe duration. Hence, it blocks this operation and informs the user of this issue, as shown in the Fig.~\ref{fig:blAttLog}. When the next two requests arrive, the coprocessor decides that both the requests are not safe and warns the user of a possible attack. In addition, the coprocessor should inform the main processor that all these treatments were blocked.

\subsubsection{Scenario 2: Bolus infusion without taking a meal\label{sec:scenario2}}

In contrast to the 1st scenario, here we are focusing on a scenario where the patient skips a meal after administering a bolus dose. Although this is usually occurs due to a patient error, an attacker also can utilize it for malicious motives.

As mentioned in section \ref{sec:scenario1}, a user administers a bolus dose when a meal is expected soon. However, it is possible that the patient misses out or delays a scheduled meal after infusing bolus insulin. If no additional carbohydrate is present after the infusion of a bolus insulin dose, the excessive insulin can cause hypoglycemia.

In addition, an attacker can cause a malicious bolus infusion.
As explained in section \ref{sec:scenario1} or as demonstrated in~\cite{Li-Healthcom11}, it is possible to make the insulin pump to infuse bolus insulin by counterfeiting a control command.
If this attack is launched when the user has no plan to take a meal or unable to respond (e.g., when asleep), this attack can successfully harm the patient.
Furthermore, as this does not violate any rule and there is no means to identify whether the user has consumed carbohydrate, it is difficult to distinguish an unsafe bolus insulin infusion and safe one at the time of infusion.
For example, a bolus insulin infusion is legal at the time of infusion when the patient originally had a plan to take a meal, but it later turns out to be unsafe when the meal is missed or delayed.
Malicious bolus insulin infusion can be partially detected beforehand by comparing the infusion command with previous infusion pattern (e.g., time and  amount)~\cite{Hei-Infocom13}.
However, if a valid amount is infused at a right time when the patient intentionally skipped a meal, a pattern-based decision may not detect it beforehand.
For an unsafe operation that cannot be identified beforehand, the next line of defense is to minimize the harm as soon as possible by alerting the patient of the excessive insulin.
Insulin overdose is typically treated by taking a skipped meal, sweets, glucose tablets, etc., if it is noticed early enough.

\begin{figure*}
\centering \includegraphics[width=\hsize]{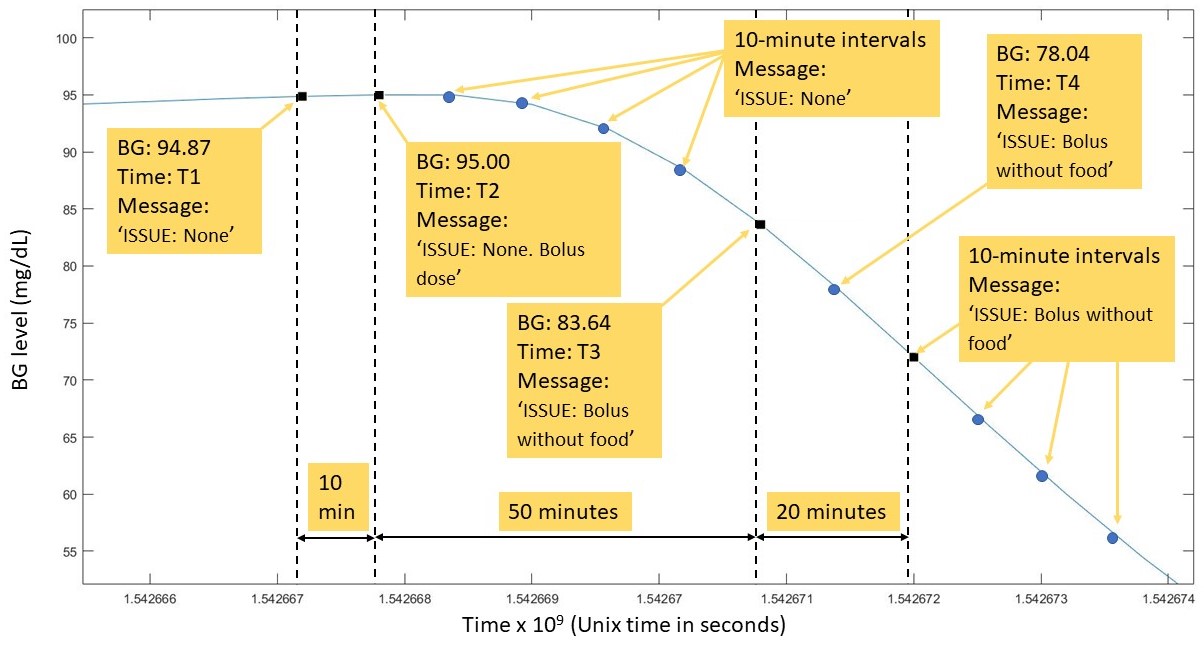}
\caption{Behavior of BG levels when bolus insulin is infused without carbohydrate.\label{fig:noFoodBolus}}
\end{figure*}

As a defense against the unsafe bolus insulin infusion, including the aforementioned cases, we define a simple time-triggered physiological rule.
Relying on that bolus insulin infusion should be followed by additional carbohydrate intake, this rule checks if an increase in the BG level is observed certain period of time after a bolus insulin dose is infused.
Whenever a bolus insulin dose is infused, an instance of this time-triggered rule is registered and the BG level is checked periodically.
In this experiment, the rule checks if the BG level is decreasing by more than 10~mg/dL after any amount of bolus insulin infusion. 
For the purpose of this experiment, these values are empirically determined from the simulation results of the GIM.
(In practice, they should be determined by the medical professionals considering the physiological characteristics of the patient.)

This simple rule is effective enough to detect the aforementioned unsafe bolus insulin infusion.
The curve in Fig.~\ref{fig:noFoodBolus} shows the BG level variation in this particular scenario. The labels shown depict the following information: BG level, the time and the message logged by the coprocessor at that time.
The patient takes the bolus insulin dose at time T2, when the BG level is 95 mg/dL. If the patient had eaten as planned, then the BG level should rise within an hour. However, it keeps decreasing as shown in the figure. Once it falls below 10 mg/dL from the previous BG level, the coprocessor decides that the patient has not taken any food and starts warning the patient at time T3. It keeps warning the patient until the BG level starts to rise.  
This prevents a severe hypoglycemia by detecting unsafe operation in this early stage.

Note that this physiological rule does not aim any specific human error or security attack as the source of the unsafe operation.
By enforcing a simple rule that checks the sensor input, triggered by a communication between the main microcontroller and the actuator, it can detect the potentially harmful consequences of the unsafe operation and inform the patient.

\subsubsection{Scenario 3: Taking a meal without a bolus infusion\label{sec:scenario3}}

In this scenario, we look at what ensues when the patient eats without taking a bolus dose. 

It is entirely possible that the patient forgets or misses out administrating a bolus dose before a meal. Since basal insulin doses do not compensate sudden intakes of carbohydrates, often there is not enough insulin in the blood stream to counteract the carbohydrate intake from a meal. As a result, the patient's BG level can increase drastically and can even cause hyperglycemia incase he/she forgets a bolus dose.

Similar to the above scenario, this too does not violate any rule and there is no means for the device to identify whether the user has consumed carbohydrate. Hence, the only way the device can identify the excess is carbohydrate is when the blood sugar starts to increase in a dramatic way. 
Even then, the device will identify this as a sudden increase in BG level and try to counteract it with one or more basal insulin doses. However, since the maximum basal insulin is usually configured to a value lesser than a typical bolus insulin dose (this depends on the amount of carbohydrate the patient is planning to take), the BG level is not controlled as expected.

Hence, the alternative is to alert the patient before the BG value reaches a critical level. When the patient receives the warning, he/she can manually administer enough insulin to compensate the sudden increase in BG values. 

\begin{figure*}
\centering \includegraphics[width=\hsize]{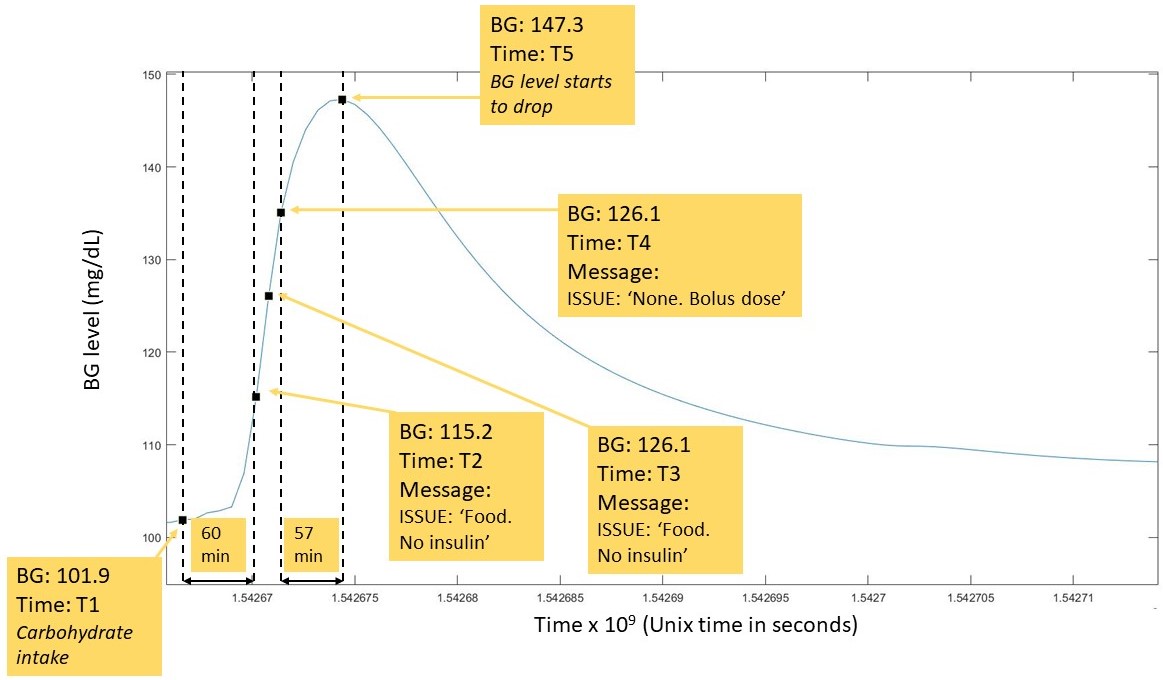}
\caption{Behavior of BG levels when carbohydrate is taken without insulin.\label{fig:noBolusFood}}
\end{figure*}

As a defense against missing bolus doses, a time-triggered physiological rule is defined.
This rule checks if there's a substantial increase in the BG level over a certain period of time. This check is triggered whenever an increase in the BG levels is observed. (e.g. increment of 15 mg/dL over a period of 1 hour without any carbohydrate intake)
In this experiment, the rule checks if the BG level has increased by more than 10~mg/dL within a period of 90 minutes.
As mentioned in section \ref{sec:scenario2}, these values are empirically determined from the simulation results of the GIM.

The results of this rule check can be seen in Fig.~\ref{fig:noBolusFood}. At time T1 the patient consumes 75g of carbohydrates without any bolus insulin. As a result the BG values rise considerably and reaches a value of 115.2 mg/dL after an hour (i.e. at T2). Since the coprocessor checks the BG levels at every 10 minutes, at this point it identifies this sudden increment and notifies the user immediately, and keeps on doing so until it registers a bolus dose infusion. After a couple more warnings are issued, the user administers a bolus dose at T4. Since the coprocessor registers this bolus intake, it stops warning the user from thereafter.  

An alternative to this would be to program the device to identify the increment in the BG values and infuse some insulin. However, since there is no meal intake registered, the device assumes that this treatment as another basal infusion. For safety purposes, the basal doses are limited by a maximum basal value, which is usually smaller than a bolus dose. In fact, whenever an increase in BG level similar to the above scenario is seen, the artificial pancreas tries to treat it using a basal dose. However, this is not enough to completely counteract the carbohydrate intake. Another possibility is to have the coprocessor programmed to infuse a pre-programmed dose of bolus insulin to mitigate the effects of BG level increase. However, since the device has no knowledge of the amount of carbohydrate taken, it can infuse an excessive or an inadequate amount of bolus insulin. Both these cases can cause undesrable effects on the patient's health.


\color{black}

\subsubsection{Scenario 4: Wireless Attack Exploiting Buffer Overflow}\label{sec:scenario4}
In this attack, an unsafe operation is triggered by exploiting a buffer overflow.
A buffer overflow is corruption of data values in memory addresses adjacent to a data buffer that occurs when the input data does not fit within the buffer.
If the input data is deliberately designed to alter program execution path, in may result in executing illegal, potentially harmful, code.
This kind of attacks can be launched against embedded systems through wireless programming~\cite{Yang-MobiHoc08, Giannetsos-CJ10}.

\begin{figure*}
\centering \includegraphics[width=\hsize]{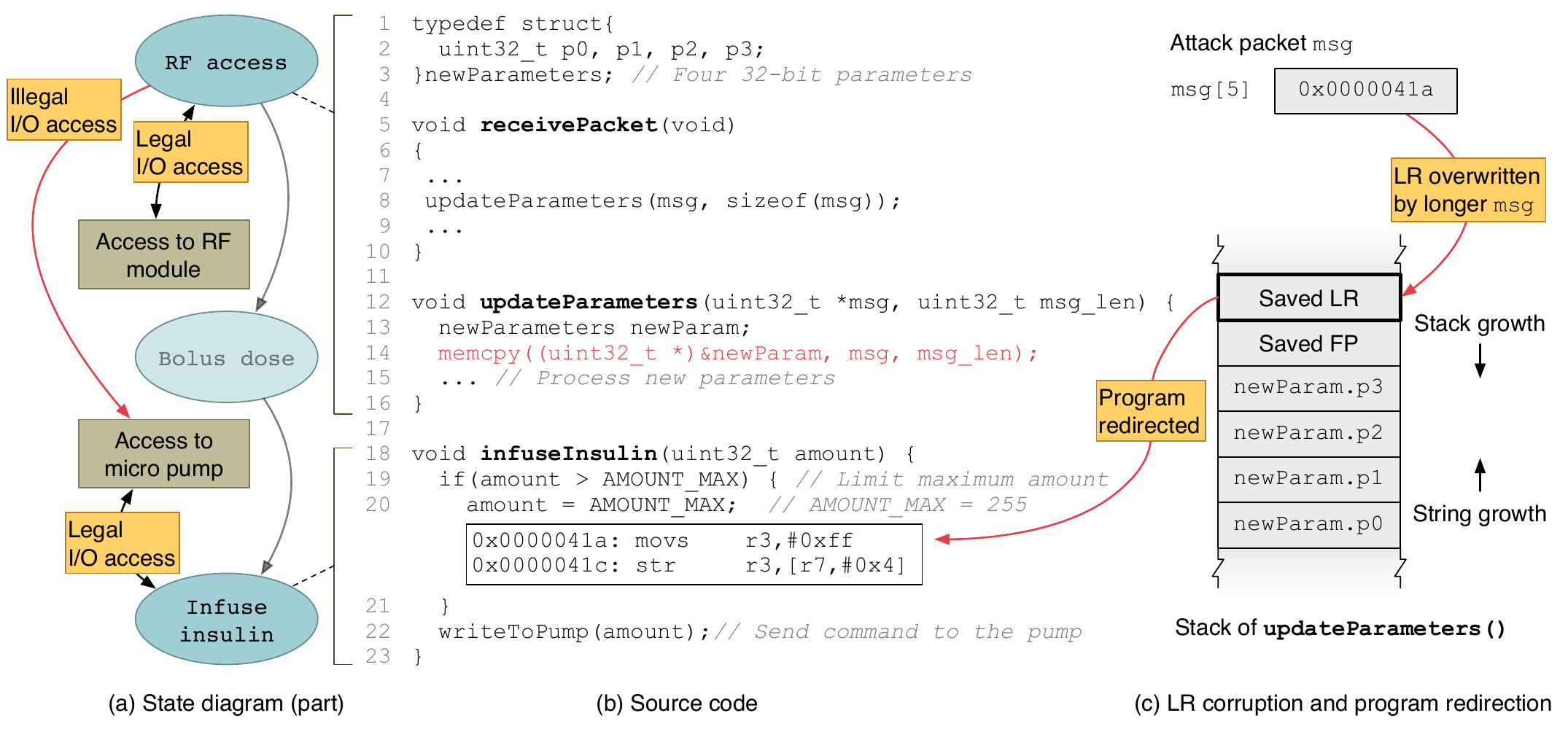}
\caption{Buffer overflow that triggers the micro pump to infuse the maximum amount of insulin.\label{fig:buffer_overflow}}
\end{figure*}

In this experiment, we exploit the parameter reprogramming feature through the wireless channel.
A snippet of the source code is listed in Fig.~\ref{fig:buffer_overflow}(b).
First, the packet \texttt{msg} received in the \texttt{receivePacket()} function is passed over to \texttt{updateParameters()} to update parameters.
We introduce a vulnerability here by using \texttt{memcpy()} without boundary checking.
As shown in Fig.~\ref{fig:buffer_overflow}(c), by passing \texttt{msg} that is longer than the buffer \texttt{newParam}, we can overwrite the link register (LR) so that the program can be redirected to any memory address.

In the \texttt{infuseInsulin()} function, the argument \texttt{amount} is checked and limited not to exceed the maximum of the data range (255 in this example).
Originally, the assignment statement of Line 20 should be executed only when \texttt{amount} is greater than \texttt{AMOUNT\_MAX}.
However, if the program is redirected to Line 20 from outside this function, it can bypass the \texttt{if} clause and unconditionally assign \texttt{AMOUNT\_MAX} to \texttt{amount}.
In this example, the address of the Line 20 is \texttt{0x0000041a}.
By overwriting this address on the LR, we can make the insulin pump immediately infuse the maximum amount of insulin.

Modern microcontrollers adopt protection mechanisms such as NX (No-eXecute) or XN (eXecute Never) bit that prohibit code execution from the stack for security purpose.
However, this attack only executes a subroutine that is already present in the original code, and does not execute code from injected input on the stack, and hence it can bypass such protection mechanisms.

An occurrence of buffer overflow is detected by the I/O access rule.
The program is in the \texttt{RF access} state until the buffer overflow takes place from \texttt{mempcy()}.
It is then redirected to \texttt{infuseInsulin()} directly, without sending program state transition messages to the safety coprocessor.
When \texttt{writeToPump()} is called, the safety coprocessor captures the anomaly that the micro pump is accessed from the \texttt{RF access} state.
The I/O access rule restricts that the micro pump can only be accessed from the \texttt{Infuse insulin} state  as shown in Fig.~\ref{fig:buffer_overflow}(a).
As a result, the micro pump driver command is blocked by the safety coprocessor, and does reach the pump driver.

\begin{figure}
\centering \includegraphics[width=10cm]{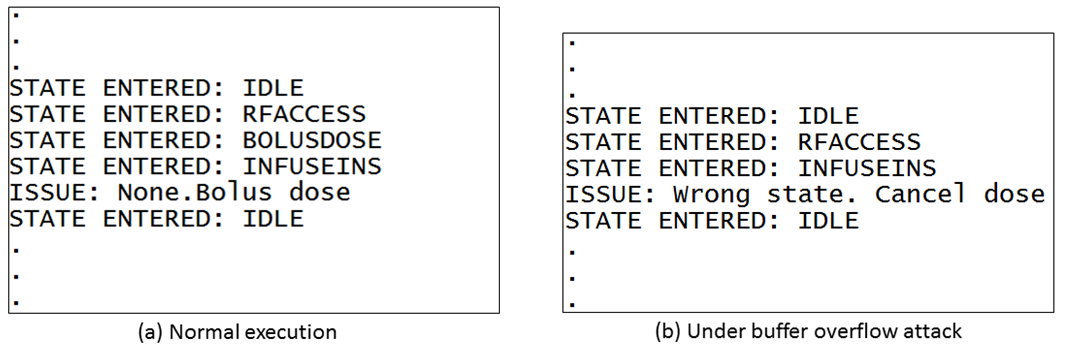}
\caption{Decoded log from the safety coprocessor.\label{fig:exp_buffer_overflow}}
\end{figure}

Logs in Fig.~\ref{fig:exp_buffer_overflow} show more details of the detection process.
These are decoded messages of the logs generated by the safety coprocessor while it enforces the safety rules.
When no buffer overflow attack is launched, the normal state transitions and I/O accesses occur as shown in Fig.~\ref{fig:exp_buffer_overflow}(a).
We can see the accesses to the RF module and the micro pump are performed only in permitted state, \texttt{RF access} and \texttt{Infuse insulin} states, respectively.
On the other hand, when a buffer overflow attack is launched as described in Section~\ref{sec:scenario4}, the micro pump is accessed from the \texttt{RFACCESS} state.
This is a violation of an I/O access rule 2, which restricts access to the micro pump (the red arrow in Fig.~\ref{fig:buffer_overflow}(a)). Hence the command is blocked by the safety coprocessor. 

Note that our I/O access rule is not defined to specifically detect buffer overflow, and so any attacks or malfunctions that manifest as similar I/O access violations will be captured by this rule.
Also, this rule is orthogonal to conventional buffer overflow detection techniques~\cite{Shao-ACSAC03, Cowan-SSYM98}.


\subsubsection{Scenario 5: Main processor stops responding/disabled}\label{sec:scenario5} In this scenario, we look at the kind of steps the coprocessor can take in case the main processor becomes non-functional.
The main processor could become unresponsive due to a number of reasons such as a power failure, an intrusion by an attacker or even an unforeseen bug in the source code. If such an event is not noticed by the patient, it can have serious repercussions on the patient's health. For instance, if an attacker successfully disables the artificial pancreas with the patient unaware, the patient's BG level may increase without control, potentially causing serious harm to the patient. 

Regardless of the cause, it is difficult to detect an unresponsive state unless the device's state is explicitly checked. Since most of these devices are designed to operate autonomously, users seldom do such a check. Hence, it is within reason to assume that a device failure as such can go unnoticed for an unspecific period of time. 

To resolve such an issue, we utilize the safety coprocessor and a time-triggered rule. The corpocessor keeps track of the messages received via the IMC channel and stores the time the last IMC message was received. If a specified time elapses without receiving any message, then the coprocessor decides that the main processor has entered an unresponsive state. 

\begin{figure}
\centering \includegraphics[width=12cm]{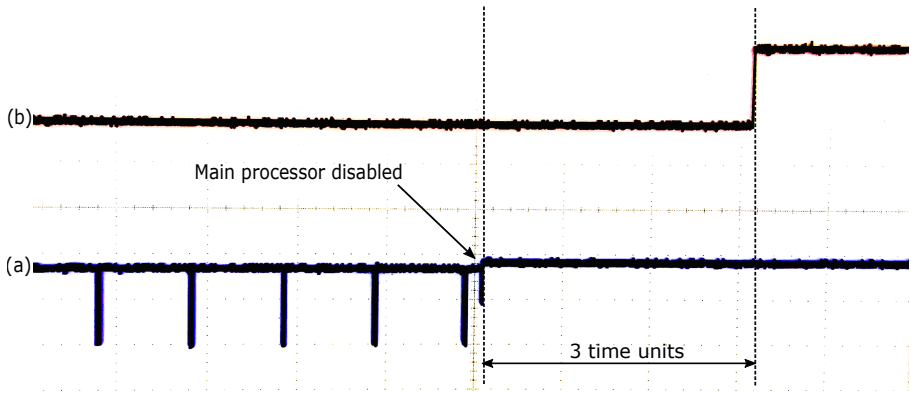}
\caption{Waveform of (a) Periodic message from the main processor and, (b) Signal set high to warn the patient.\label{fig:timeoutAction}}
\end{figure}

As shown in Fig.~\ref{fig:timeoutAction}, in this implementation we have set an interval equivalent to 3 time units as the maximum timeout for a message. Here, a time unit is the duration between two periodic messages sent by the main processor. We feel that this duration is long enough so that the coprocessor will not raise a false alarm and short enough that the absence of the main processor does not have serious repercussions on the patient. However in practice, it is better to let a physician set the duration depending on the patient's attributes rather than having a fixed value. 

Several different types of actions can be taken, depending on the disease/condition and the patient's characteristics. For a less threatening condition, the coprocessor can notify the patient regarding the failure, so that the patient can take appropriate action. For instance, the safety coprocessor that is used in the artificial pancreas platform issues a warning to the patient via a flashing LED. However, in a more critical scenario, where more diligent action is necessary, the coprocessor can be designed to take more drastic actions such as restarting the main processor or take a set of predetermined steps to protect the patient until he/she can obtain proper care. However, care must be taken when designing such countermeasures so as not to jeopardize the patient's safety.

\section{Conclusions}\label{sec:conclusions}
In this paper, we propose a HW/SW framework that improves the safety of IWMDs.
The proposed framework includes a set of safety rules and a rule-check mechanism that acts as both assessor and guardian against unsafe IWMD operations.  
The framework includes an off-the-shelf property checker and a safety coprocessor. The property checker checks if the rules are followed by the IWMD, based on the semantics of its firmware. The rules that are impossible to be asserted in this manner are left to be tested during run time, using the safety coprocessor. The low-power safety coprocessor monitors both extrinsic state and internal state of the IWMD to detect unsafe operations due to user errors, software bugs, or security attacks.
Based on high visibility and accessibility to the behaviors of the target IWMD, the safety coprocessor can enforce three types of safety rules: state transition rules, I/O access rules, and physiological rules to identify various unsafe operations.
Once it identifies an unsafe operation, it takes necessary protective actions, including raising a warning, suspending the main microcontroller, and performing emergency life-supporting operations.
We presented the design and implementation of the safety-enhanced controller board based on the proposed architecture.
We demonstrated the effectiveness of the proposed safety architecture with various safety risks using a prototype artificial pancreas system.
We believe that the proposed architecture can be widely adopted to address a variety of practical safety concerns arising from the ever-increasing connectivity and software complexity of IWMDs.


\color{black}


\bibliographystyle{ACM-Reference-Format}
\bibliography{mainFile}

\end{document}